\documentstyle[epsfig,twocolumn]{jpsj}
\title{Electronic Structure and Phase Transition in V$_2$O$_3$: Importance of
3$d$ Spin-Orbit Interaction and Lattice Distortion
}

\author{
Arata {\sc Tanaka}
}

\inst{
Department of Quantum Matter,  ADSM\\
 Hiroshima University, Higashi-Hiroshima 739-8530, Japan
}

\recdate{ }

\abst{
The 3$d$ electronic structure
 and phase transition in pure and Cr doped V$_2$O$_3$ are theoretically 
investigated in relation to the 3$d$ spin-orbit interaction and lattice distortion.
A model consisting of the nearest-neighbor V ion pair with full degeneracy of the
3$d$ orbitals  is studied within the many-body point of view.
It is shown that each V ion with $S=1$ spin state has a large orbital magnetic
moment $\sim 0.7 \mu_{\rm B}$ and no orbital ordering occurs
in the antiferromagnetic insulating (AFI) phase. The anomalous
resonant Bragg reflection found in the AFI phase is attributed to the magnetic ordering.
In the AFI and paramagnetic insulating (PI) phases, Jahn-Teller like lattice instability leads to
tilting of the V ion pairs from the corundum $c$-axis
and this causes large difference in the orbital occupation between the paramagnetic metal and the insulating phases, 
which is consistent with linear dichroic V 2$p$ XAS measurements.
 To understand the AFI to PI
transition, a model spin
Hamiltonian is also proposed. The transition is found to be 
simultaneous order-disorder transition of the 
magnetic moments and tilting directions of the
V ion pairs. Softening of elastic constant $C_{44}$ and abrupt change
in short range spin correlations observed at the transition are also explained.
}
\kword{
V$_2$O$_3$, electronic structure, phase transition, lattice distortion, 
orbital ordering, spin-orbit interaction, 
orbital moment, resonant x-ray scattering, V 2$p$ XAS
}

\begin{document}
\maketitle
\section{Introduction}
The metal-insulator (Mott) transition in V$_2$O$_3$ have been extensively
studied ever since Mott suggested the possibility of the 
localization of electrons driven by Coulomb repulsion\cite{Mott}.
Above the transition temperature $T_{\rm N}\sim 155$K, the material is paramagnetic metal (PM)
and has the corundum structure with the trigonal symmetry (space group $R\overline{3}c$).
Below $T_{\rm N}$, it is an antiferromagnetic insulator (AFI)  and 
the lattice distortion to the monoclinic structure ($I2/a$) takes place\cite{Dernier}.
Cr doped alloys (V$_{1-x}$Cr$_x$)$_2$O$_3$  
 exhibit also interesting features\cite{McWhan,Jayaraman,Menth,McWhan2,Kuwamoto}.
For $0.005<x<0.018$,
at a temperature above the AFI $\to$ PM transition, the material further
undergoes a PM to paramagnetic insulator (PI) phase transition
with no change in the crystal structure.
For much higher Cr concentration $x>0.018$, 
the PM phase disappears and a AFI $\to$ PI transition takes place.

Although V$_2$O$_3$ and its alloys are often referred to a typical example of the 
Mott-Hubbard system,
there are experimental facts showing the complex aspects of these materials,
which can not be explained by the simple single-band Hubbard model.
In the AFI phase, they exhibit unusual magnetic order, which breaks
corundum trigonal symmetry\cite{Moon}.
Neutron scattering experiments have been performed by Bao {\it et al.}
and abrupt change in spatial spin correlation functions has been reported both
at the AFI $\to$ PM and the AFI $\to$ PI transitions\cite{Bao,Bao2}.
This switching of spin correlation
indicates that the phase transitions can not be considered
as usual order-disorder magnetic transition.
Softening of the $C_{44}$  elastic constant
has been observed in ultrasonic wave\cite{Yang} and neutron scattering\cite{Yethiraj}  experiments at the AFI $\to$ PI transition, 
indicating nearly second order feature of this phase transition as well as
strong electron-lattice coupling.

Even now, the electronic structure of these materials are still controversial problem and
above experimental facts suggest that
much realistic model including the orbital degeneracy of the 3$d$ orbital and the lattice
distortion is necessary to explain them.
In V$_2$O$_3$, the $O_h$ crystal field  splits the  3$d$ orbitals of each V ion 
 into two fold  $e_g$ and three fold $t_{2g}$ orbitals
and the $t_{2g}$ orbitals are further separated into low laying two fold  $e^\pi$ and
non-degenerate $a_1$ orbitals by small trigonal field.
In localization limit, two 3$d$ electrons in the V$^{3+}$ ion are
accommodated in the lowest $e^\pi$ orbitals with $S=1$ spin state.
Castellani {\it et al.} have proposed a model including 
these $e^\pi$ and $a_1$ orbitals\cite{Castellani}.
In the corundum structure, two V ions forms pairs along the $c$-axis and
in their model, the $a_1$ orbitals on two V ions of the pair
forms  molecular orbitals. The bonding $a_1$ orbital accommodates two electrons out of 
four in the pair and remaining two electrons occupies one of
$e^\pi$ orbital in each V ions, resulting in $e^\pi a_1$
configuration with $S=1/2$ spin state in each V ion.
They also propose a long-range ordering of orbital 
occupation in the two fold $e^\pi$ orbital to explain the complex
antiferromagnetic ordering in the AFI phase.

However, recent measurements on the linear dichroism 
of V 2$p$ x-ray absorption spectra (XAS) in V$_2$O$_3$ have revealed that
the 3$d$ electron configuration is not $S=1/2$ $e^\pi a_1$
but mixture of $e^\pi e^\pi$ and $e^\pi a_1$
with $S=1$ Hund's rule spin state\cite{Park}.
In addition, large differences in  $e^\pi e^\pi$ to $e^\pi a_1$ ratio
have been found among the AFI, PI and PM phases.
Indeed, such $S=1$ Hund's rule spin state 
is also obtained in recent ``LDA+$U$''\cite{Anisimov,Anisimov2} band structure calculation\cite{Ezhov},
in which a  $S=1$ $e^\pi e^\pi$ ground state
having  experimentally observed antiferromagnetic order 
with no orbital degeneracy 
is favorable for the low temperature monoclinic phase.
Mila {\it et al.} have proposed a model 
with a large intraatomic exchange interaction\cite{Mila,Shiina}.
In this model, spin state in each V ion remains
being $S=1$ even with strong hybridization 
between $a_1$ orbitals. The electronic state of the V ion pair is
described  as a superposition of two configurations with equal weight:
  $e^\pi a_1;e^\pi e^\pi$ (the $e^\pi a_1$ configuration in one of V ion and 
the  $e^\pi e^\pi$  configuration in the other) and $e^\pi e^\pi;e^\pi a_1$.
In this model, two fold orbital degeneracy is present in this V$_2$ molecular
orbital and the complex antiferromagnetic order in the AFI phase is stabilized 
with a ferro-type order of this V$_2$ molecular orbital.

Paolasini {\it et al.} have performed resonant x-ray scattering experiments (RXS) at the V $K$-edge
on V$_2$O$_3$ and they insisted that resonant Bragg peaks observed  in the AFI phase
are due to the 3$d$ orbital ordering\cite{Paolasini,Paolasini2}.
However, Lovesey and Knight have argued that the 
resonant Bragg peaks arise from the ordered orbital magnetic moments of the V ions\cite{Lovesey}.
Indeed, nonresonant magnetic scattering 
has been also measured and a large contribution from the orbital magnetic moment to
the total magnetic moment $M_{L}/M_{S}\sim -0.3$ was found\cite{Paolasini}.
It is thus important to know how the 3$d$ spin-orbit interaction
affects the 3$d$ electronic structure of V$_2$O$_3$.
Since this interaction couples spin and orbital degrees of freedom, 
in the presence of a large orbital angular momentum, no orbital ordering is expected 
in the AFI phase.
In above mentioned theories on the electronic structure
of V$_2$O$_3$, however, the effects of the 3$d$ spin-orbit interaction in the V ions is
presumed to be small. 

The purpose of this paper is to clarify the effects of the 3$d$ spin-orbit
interaction and  the lattice distortion on the 3$d$ electronic structure and the phase transitions 
of the pure and Cr doped V$_2$O$_3$.
To this end, first of all, a V ion pair model similar to that proposed by Mila {\it et al.}
will be  considered taking into account the 3$d$ spin-orbit interaction in \S~2.
With the 3$d$ spin-orbit interaction, no orbital ordering is expected in the AFI phase and 
instead of this, a large orbital magnetic moment $\sim 0.7\mu_{\rm B}$ is
induced in the ground state. 
The relation between the monoclinic distortion and the 
magnetic anisotropy and difference in the configuration of the 3$d$ electrons
between the AFI and PM phases will be also discussed.
Second, interaction between the electronic state of the V ion pair
 and local lattice distortion caused by displacement of the pair  will be discussed in \S~3.
It will be shown that Jahn-Teller like lattice instability
occurs when the energies of the $e^\pi a_1;e^\pi e^\pi$ ($e^\pi e^\pi;e^\pi a_1$)
and $e^\pi e^\pi;e^\pi e^\pi$ states are nearly degenerate.
Third, relation between 
the magnetic ordering and the lattice distortion in the AFI phase and also 
the mechanism of the AFI $\to$ PI transition will be discussed in \S~4. 
For this purpose, an effective Hamiltonian  with exchange
interaction between the V ion pairs including lattice distortion effects is proposed.
Softening of the elastic constant $C_{44}$  and abrupt change in short range spin 
correlation functions observed at the AFI $\to$ PI transition are  reproduced with the model.
Finally,  to confirm the validity of the present model,
recent experiments on RXS
and the linear dichroic V 2$p$ XAS spectra are analyzed in \S~5 and \S~6, respectively.
The experimental results are well explained within the present model.

\section{V ion Pair Model \label{Model}}
\subsection{Model Hamiltonian\label{model}}

\begin{figure}
\begin{center}
%\figureheight{3cm}
\epsfig{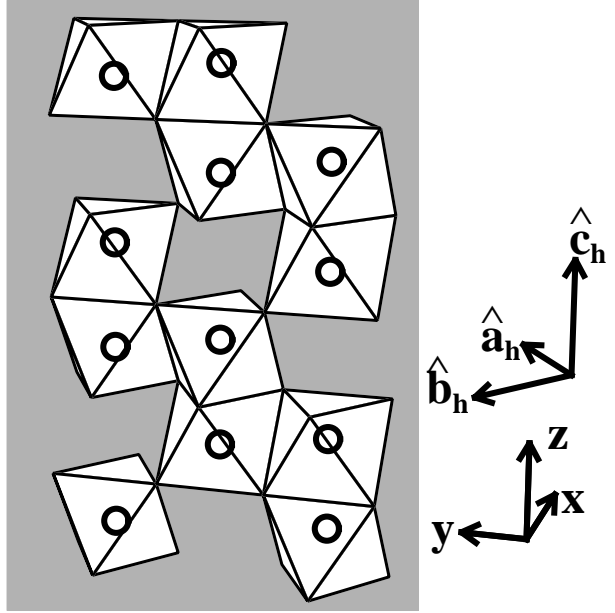}
\end{center}
\caption{Corundum structure of V$_2$O$_3$.
 Each V ion (open circles) is octahedrally surrounded by six oxygen ions (octahedra).
 Along the $\mib{c}_h$-axis, the V ion pairs surrounded by the face-sharing two oxygen-octahedra are seen
 whereas in $\mib{a}_h$-$\mib{b}_h$ plane, each V ion has three neighbors with 
 edge-sharing  oxygen-octahedra.
}\label{cry}
\end{figure}
Each V ion in the corundum structure of V$_2$O$_3$ is
octahedrally surrounded by six oxygen ions (see Fig.~\ref{cry}).  
An octahedral crystal field and hybridization between the V 3$d$ and O 2$p$ orbitals
split the energy level of the 3$d$ orbital  into the $e_g$ and $t_{2g}$ levels and the low laying
$t_{2g}$ orbitals accommodate two electrons in each V$^{3+}$ ion.
A small trigonal distortion of the oxygen octahedron
 along the $\mib{c}_h$-axis further splits the
$t_{2g}$  level  into a doubly degenerate $e^\pi$ and a non-degenerate $a_1$ levels 
with the $C_{3v}$ symmetry. 
In the corundum structure, the layers of the honeycomb lattice of V ions 
are stacking along the $\mib{c}_h$-axis and each V ion in the layers
forms a V ion pair along the $\mib{c}_h$-axis with a V ion in the adjacent layer.
While the $a_{1}$ orbitals point to the direction of the $\mib{c}_h$-axis, forming
strong $\sigma$ bonds in the V ion pairs,
 the $e^\pi$ orbitals extend their lobes toward the direction perpendicular to
the $\mib{c}_h$-axis,  having $\pi$ bonds with the three neighbors in the layer.

As was first suggested by Allen, here, it is assumed that the electrons are delocalized 
within the nearest-neighbor V ion pair and interaction between much father V ions is week\cite{Allen}.
The two V ions of the pair along the $\mib{c}_h$-axis and electron 
hopping between them were considered based on configuration interaction approach\cite{Tanaka,Eskes,Tanaka2}.
In each V ion site, the fivefold 3$d$ orbitals are fully included 
and the Coulomb and exchange interactions among 3$d$ electrons,
the 3$d$ spin-orbit interaction and the trigonal crystal field are considered\cite{Sugano}. 
The Hamiltonian for the V ion pair model then would be written as
\begin{eqnarray}
 H_{\rm elec}&=& \sum_{i,\nu}
       \varepsilon_d(\nu) n^{~}_{i\nu}
      +\frac{U_{dd}}{2}\sum_{i,\nu,\nu'} 
        n^{~}_{i\nu} n^{~}_{i\nu'} \nonumber\\
    &+&\sum_{\nu,\nu'}
        V(\nu,\nu')(a^\dagger _{A,\nu} a^{~}_{B,\nu'}
        +a^\dagger _{B,\nu} a^{~}_{A,\nu'})\nonumber\\
     &+&\sum _{ i,\nu_1,\dots,\nu_4}
     g(\nu_1,\nu_2;\nu_3,\nu_4)
      a^\dagger _{i\nu_1}a^\dagger _{i\nu_2}
      a^{~}_{i\nu_4}a^{~}_{i\nu_3}\nonumber\\
     &+& \zeta _{d} \sum_{i,\nu,\nu'} 
                \langle \nu | \mib{l}\cdot \mib{s} |\nu' \rangle
               a^\dagger_{i\nu} a^{~}_{i\nu'}\nonumber\\
     &+&\sum_{i,\nu\neq\nu'}C_i(\nu,\nu')a^\dagger _{i,\nu} a^{~}_{i,\nu'},\label{Ham}
\end{eqnarray}
where the first and second  terms are those for  
the one body energies of the 3$d$ orbitals, the effective Coulomb repulsion energy
between the 3$d$ electrons, respectively, the third term represents the electron hopping 
between two V ion sites A and B in the V ion pair given in terms of the Slater-Koster
parameters $dd\sigma$ and $dd\pi$, 
the forth term is for the multipole interaction between the 3$d$ electrons, which is described by the Slater integrals  $F_{dd}^2$ and $F_{dd}^4$ of the 3$d$ orbital and 
the fifth term denotes the 3$d$ spin-orbit interaction
 with the coupling constant $\zeta_{d}$. The last term represents crystal field due to the monoclinic
 lattice distortion in the AFI phase.
Here, $\nu$ represents
the combined index of the species of the 3$d$ orbital
and the spin quantum number of the electron, $i=A,B$ stands for the V ion site and
$a^\dagger_{i\nu}$ creates
an electron 
on the 3$d$ orbital with $\nu$ at the V ion site $i$.

In a $C_{3v}$ crystal field, the five degenerate 3$d$ orbitals
split into one non-degenerate $a_1$ orbital
and two doubly-degenerate $e$ orbitals.
Using coordinates where $z$-axis is parallel to the $\mib{c}_h$-axis of the
corundum structure and $y$-axis to one of the three V-V bond direction in the
 $\mib{a}_h$-$\mib{b}_h$ 
basal plane (see Fig.~\ref{cry}), the wavefunctions for the 3$d$ orbitals
 in terms of the $C_{3v}$ basis functions are written as
\begin{eqnarray*}
    |a_{1}:i\rangle &=& |3z^2-r^2\rangle, \\
   |e_u^\pi:i\rangle&=&
            s_i\sqrt{\frac{2}{3}} |x^2-y^2\rangle  -\sqrt{\frac{1}{3}} |zx\rangle,\\
   |e_v^\pi:i\rangle&=&
          -s_i\sqrt{\frac{2}{3}} |xy\rangle            -\sqrt{\frac{1}{3}} |yz\rangle,\\ 
   |e_u^\sigma:i\rangle&=&
            s_i\sqrt{\frac{1}{3}} |x^2-y^2\rangle  +\sqrt{\frac{2}{3}} |zx\rangle,\\
   |e_v^\sigma:i\rangle&=&
          -s_i\sqrt{\frac{1}{3}} |xy\rangle             +\sqrt{\frac{2}{3}} |yz\rangle,
\end{eqnarray*}
where the two kinds of $e$ orbitals in the $C_{3v}$ symmetry
originating from the $t_{2g}$  ($e_g$) orbitals
in the $O_h$ symmetry is specified by superscript $\pi$ ($\sigma$),
the subscripts $u$ and $v$ indicate two different kinds of 
basis functions in each $e$ symmetry orbitals
 and $s_i$ takes either $+1$ or $-1$ depending on the orientation
 of the oxygen octahedron surrounding the $i$-th V ion site and 
 has different sign between the two V ions in each V ion pair.
 The energies of these orbitals $\varepsilon_d(\nu)$ can be expressed
 in terms of the octahedral $Dq$
and the trigonal $D_{\rm trg}$ crystal fields parameters:
\begin{eqnarray*}
\varepsilon_d(a_{1}) &=&-4Dq+2D_{\rm trg}/3,\\
\varepsilon_d(e_u^\pi) &=&\varepsilon_d(e_v^\pi)=-4Dq-D_{\rm trg}/3,\\
\varepsilon_d(e_u^\sigma)&=&\varepsilon_d(e_v^\sigma)=6Dq.
\end{eqnarray*}
\begin{figure}
\begin{center}
%\figureheight{3cm}
\epsfig{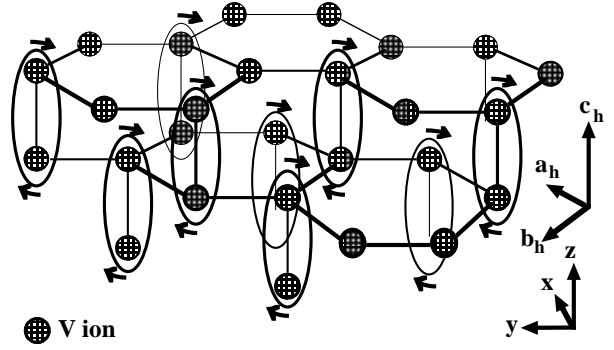}
\end{center}\par
\caption{
Schematic drawing of the monoclinic distortion of  
V$_2$O$_3$ in the AFI phase. The circles denote V ions and
the ovals show the V ion pairs. The allows represent the 
directions of the tilting of the V ion pairs in the 
AFI phase.
}\label{tilting}
\end{figure}

In the AFI phase, the crystal structure is distorted from the corundum structure
to the monoclinic one.
This distortion causes tilting of the V-V bond direction of every V ion pair from the 
$z$-axis toward the negative side of the $y$-axis
and  breaks three fold rotational symmetry (see Fig.~\ref{tilting}).
A low symmetry crystal field, which hybridizes the $e_v$ and $a_1$ 
orbitals and also the $e_v$ and $e_u$ orbitals on the same site, is present in this phase.
Although there are five independent parameters for this low symmetry
crystal field $C_i(e_v^\pi,a_1)$,
$C_i(e_u^\pi,e_v^\pi)$, $C_i(e_v^\sigma,a_1)$, $C_i(e_v^\sigma,e_u^\sigma)$ and
$C_i(e_v^\pi,e_u^\sigma)$, 
the most important are those related to the $e^\pi$ orbitals: $C_i(e_v^\pi,a_1)$
and $C_i(e_u^\pi,e_v^\pi)$. Other three parameters involving $e^\sigma$ orbitals
are, for simplicity, all assumed to be zero. Note that there are relations
between these parameters for the two V ion sites in the pair: 
$C_{\rm A}(e_v^\pi,a_1)=C_{\rm B}(e_v^\pi,a_1)$, 
$C_{\rm A}(e_u^\pi,e_v^\pi)=-C_{\rm B}(e_u^\pi,e_v^\pi)$.

From recent analysis of high-energy spectroscopy spectra, 
V$_2$O$_3$ is not simply ascribed to the Mott-Hubbard type insulator, where
the 3$d$ orbital is the main component of the valence-band top and the 
conduction-band bottom,
but intermediate between the Mott-Hubbard and charge-transfer type
insulators, where both the valence-band top and conduction-band bottom consist of
highly mixed states of the 3$d$ and O 2$p$ orbitals\cite{Bocquet,Uozumi,Uozumi2}.
It should be emphasized that although the model we assume here
 does not contain the O 2$p$ orbitals explicitly,  the effects of the hybridization
 between the 3$d$ and 2$p$ orbitals can be  partially included by using renormalized 
 parameters.
 As will be discussed next section, the on-site effective Coulomb energy between the 3$d$ electrons should be strongly renormalized by this $d$-$p$ hybridization.

\subsection{Effect of $d$-$p$ hybridization\label{dp} }

Before discussing the details of the ground state properties
 obtained from the model, in this subsection, we will briefly consider
 the effects of hybridization between the V 3$d$ and the nearest neighbor O 2$p$ orbitals.
 To this end,  here we compare a single V$^{3+}$ ion model
 with a VO$_6$ cluster model\cite{Tanaka}, in which the V 3$d$ orbital and 2$p$ 
orbitals of the six oxygens surrounding the V ion octahedrally  are included.
The parameters  adopted in the cluster model are the effective on-site 3$d$-3$d$ Coulomb energy $U_{dd}$=4.0~eV, 
  the charge transfer energy between the 3$d$ and 2$p$ orbitals $\Delta=4.0$~eV,
 the Slater-Koster parameters for the 
 3$d$-2$p$ hybridization $pd\sigma=-1.73$~eV and $pd\pi=0.75$~eV and
 the octahedral crystal field $10Dq=1.0$~eV.
 These parameter values are deduced from  the high-energy spectroscopy experiments\cite{Bocquet}.
 The configurations $3d^{2+n}\underline{L}^n$ ($n=0,1,2,3$) are assumed 
 for the ${{\rm VO}_6}^{9-}$ cluster, where $\underline{L}$ denotes a hole in the O 2$p$ orbital.
  For the Slater integrals of the 3$d$-3$d$ multipole interactions $F^2_{dd}$ and $F^4_{dd}$,
 the atomic Hartree-Fock values
 reduced to 80\% are used for the VO$_6$ cluster model. 
\begin{table}
\caption{Eigen energies relative to the ground state energy
for  the lowest ten states calculated with the VO$_6^{9-}$ cluster and the 
V$^{3+}$ ion models are shown. The representation in the $O_h$ symmetry and
 dominant electron occupation in the $3d^2$ configuration of each eigen state 
 (enclosed by the parentheses) are also listed.
\label{levels}}
\begin{tabular}{llcc}\hline
\multicolumn{2}{c}{Eigen state} &\multicolumn{2}{c}{ Energy (eV)} \\
        &            & Cluster & Ion  \\ \hline
$^3T_1$ &$(t_{2g}^2)$   &  0   & 0 \\
$^1T_2$ &$(t_{2g}^2) $  & 1.40 & 1.37\\
$^1E~$  &$(t_{2g}^2 )$  & 1.42 & 1.40 \\
$^3T_2$ &$(t_{2g}e_g )$ & 1.70 & 1.53 \\
$^3T_1$ &$(t_{2g}e_g )$ & 2.68 & 2.82 \\
$^1A_1$ &$(t_{2g}^2 )$  & 2.81 & 2.84 \\
$^1T_2$ &$(t_{2g}e_g )$ & 3.14 & 3.10 \\
$^1T_1$ &$(t_{2g}e_g )$ & 3.39 & 3.23 \\
$^3A_2$ &$(e_g^2)$      & 3.57 & 3.42 \\
$^1E~$  &$(e_g^2)$      & 4.86 & 4.77 \\ \hline
\end{tabular}
\end{table}

 In Table \ref{levels}, the lowest ten eigen energies with respect to the ground state energy
 calculated with the ${{\rm VO}_6}^{9-}$ cluster and the single V$^{3+}$ ion models are listed.
 For simplicity, the 3$d$ spin-orbit interaction is not considered in 
 both the models. For the V$^{3+}$ ion model, the Slater integrals $\tilde{F}^{2,4}_{dd}$
 reduced to 90\% of those for the cluster model
  and $10\tilde{Dq}=1.7$~eV are  adopted for the V$^{3+}$ ion
 model.
In the ground state of the ${{\rm VO}_6}^{9-}$ cluster model, the weights of the configurations 
$3d^2$, $3d^3\underline{L}$,  $3d^4\underline{L}^2$ 
and  $3d^5\underline{L}^3$ are 41.8\%, 44.6\%, 12.5\% and 1.1\%, respectively.
Although such strong 3$d$-2$p$ hybridization, it is clearly seen 
in Table \ref{levels} that 
 the energy level scheme of the ${{\rm VO}_6}^{9-}$ cluster 
is well  reproduced by the V$^{3+}$ ion model with the renormalized parameters 
$\tilde{F}^{2,4}_{dd}$ and $10\tilde{Dq}$.
Similar low laying level correspondence was found between the ${{\rm VO}_6}^{10-}$ cluster 
and  V$^{2+}$ (3$d^1$) ion calculations and also between the ${{\rm VO}_6}^{8-}$ cluster
and  V$^{4+}$ (3$d^3$) ion calculations.
Because of this correspondence of the energy levels of the two, 
the V ion model would be good starting point
to describe the electronic state of V$_2$O$_3$.

The effective on-site Coulomb energy between ``3$d$ electrons'' for the V ion model is 
greatly reduced compared to the model with the O 2$p$ orbitals.
The on-site effective Coulomb energy 
between two different ``$t_{2g}$ orbitals'' $\tilde{U}'$ can be estimated
from the ${{\rm VO}_6}$ cluster model by
$\tilde{U}'=E_g({{\rm VO}_6}^{8-})+E_g({{\rm VO}_6}^{10-})-2E_g({{\rm VO}_6}^{9-})$,
 where 
$E_g$'s denote the ground state energies
of the clusters with different charges.
The on-site Coulomb energy between electrons on the same ``$t_{2g}$ orbital'' $\tilde{U}$
and the on-site exchange energy of the  ``$t_{2g}$ orbitals'' $\tilde{J}$ was obtained
by using $\tilde{U}=\tilde{U}'+6\tilde{F}_2+40\tilde{F}_4$ 
and $\tilde{J}=3\tilde{F}_2+20\tilde{F}_4$\cite{Sugano}.
The values obtained are $\tilde{U}=3.61$~eV, $\tilde{U}'=2.31$~eV and $\tilde{J}=0.65$~eV.
Similarly, the renormalized value for $U_{dd}$ was obtained by $\tilde{U}_{dd}$
=$\tilde{U}'+\frac{4}{9}\tilde{F}_2-10\tilde{F}_4$=2.07~eV.
These are the values which should be used in models in which
 the O 2$p$ orbitals are not explicitly included.
For comparison, these quantities for the bare $t_{2g}$ orbitals obtained
with $U_{dd}=4.0$~eV and $F_{2,4}$ reduced to 80\% from Hartree-Fock values
are $U=5.60$~eV, $U'=4.15$~eV and $J=0.73$~eV.
\begin{table}
\caption{Values of the renormalized parameters adopted in the V ion pair model.
For the $d$-$d$ hopping integrals and the crystal field parameters $C_{\rm A}(e_v^\pi,a_1)$
and $C_{\rm A}(e_u^\pi,e_v^\pi)$, values for the AFI and PM phases are indicated.
All values are units of eV. \label{param}}
\begin{tabular}{ccccc}\hline
           &   &             & (AFI)       & (PM)     \\ 
$U_{dd}$    &2.07&   $dd\sigma$  &$-0.735$ &  $-0.793$\\
$F^2_{dd}$ &7.29&  $dd\pi$  & ~~0.100 & ~~0.109  \\
$F^4_{dd}$ &4.57&   $dd\delta$ & 0 & 0 \\
$10Dq$                     & 1.7&  $C_{\rm A}(e_v^\pi,a_1)$         & ~~$0.02 $   & 0\\
$\zeta_d$ &0.023& $C_{\rm A}(e_u^\pi,e_v^\pi)$  & $- 0.02$  & 0\\ \\ \hline
\end{tabular}
\end{table}

Because of this strong $d$-$p$ mixing, the hopping integrals $dd\sigma$, $dd\pi$ and $dd\delta$
between the two V ion sites in the pair must be also renormalized.
Effective $d$-$d$ hopping matrix elements, where
the influence of the $d$-$p$ mixing  is considered 
within the second-order perturbation theory, are
\begin{eqnarray*}
V(a_1,a_1)&\simeq&\frac{1}{N_\pi}\Bigl(dd\sigma-\frac{2(pd\pi)^2}{3\Delta} \Bigr),\\
V(e_\gamma^\pi,e_\gamma^\pi)&\simeq&\frac{1}{N_\pi}
    \Bigl(-\frac{1}{3}dd\pi+\frac{2}{3}dd\delta
          + \frac{4(pd\pi)^2}{3\Delta}  \Bigr),\\
V(e_\gamma^\sigma,e_\gamma^\sigma)&\simeq&\frac{1}{N_\sigma}
    \Bigl(-\frac{2}{3}dd\pi+\frac{1}{3}dd\delta
          +\frac{(pd\sigma)^2}{2\Delta}  \Bigl),\\
V(e_\gamma^\sigma,e_\gamma^\pi)&=&V(e_\gamma^\pi,e_\gamma^\sigma)\\
&\simeq&\frac{\sqrt{2}}{3\sqrt{N_\pi N_\sigma}}\Bigl(dd\pi+dd\delta
         -\frac{\sqrt{6}(pd\pi)(pd\sigma)}{\Delta} \Bigr),
\end{eqnarray*}
where $\gamma=u,v$,  $N_\pi=1+4(pd\pi)^2/\Delta$ and $N_\sigma=1+6(pd\sigma)^2/\Delta$.
Using parameter values for $pd\sigma$, $pd\pi$ and $\Delta$ deduced from the
 high-energy spectroscopy experiments mentioned before,  it is found that
values of the matrices other than $V(a_1,a_1)$  is largely reduced from those without
$d$-$p$ hybridization.
Such the effects of the $d$-$p$ hybridization was included approximately  by reducing 
the value of $dd\pi$ to 25\% from the bare value.
Harrison's values are adopted as for the bare values of the $d$-$d$ hopping integrals\cite{Harrison}.
The renormalized parameters adopted in the V ion pair model are summarized in Table. \ref{param}.

\subsection{Properties of ground state\label{GS} }
 
\begin{figure}
\begin{center}
%\figureheight{3cm}
\epsfig{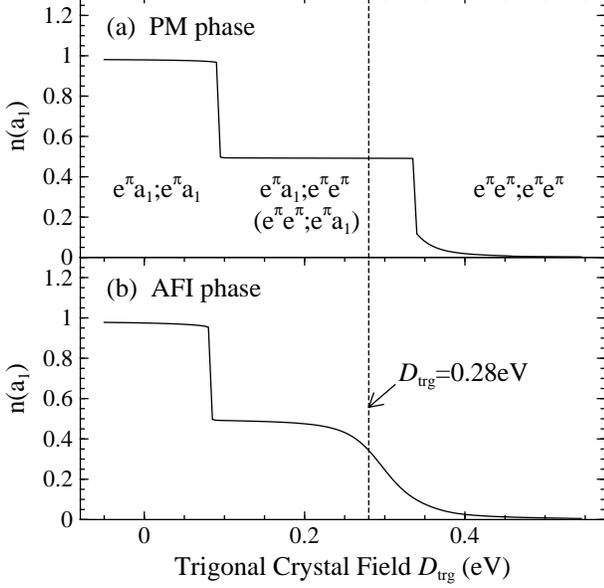}
\end{center}\par
\caption{
 Trigonal crystal field $D_{\rm trg}$ dependence of 
the  $a_1$ orbital occupation number per V ion in the ground state 
for the PM (a) and AFI (b) phases are shown.
The vertical line is positioned at $D_{\rm trg}=0.28$~eV where the ground state
gives the ratio $e^\pi a_1:e^\pi e^\pi=1:2$ deduced from the XAS experiment for the AFI phase.
}\label{a1occup}
\end{figure}
In this subsection, we will discuss the ground state properties of the V$_2$O$_3$,
using the model described in \S~\ref{model}.
In Fig.~\ref{a1occup}(a), the trigonal crystal field $D_{\rm trg}$ dependence of 
the  $a_1$ orbital occupation number for a V ion in the ground state 
for the PM phase is shown.
Two discontinuous decrease in the occupation number 
at $D_{\rm trg}=0.09$~eV and $D_{\rm trg}=0.34$~eV are seen as  $D_{\rm trg}$  increases.
Because of the Hund's rule coupling, the spins of two electrons
in each V ion site form $S=1$ spin state in the ground state.
This is very different result to the model proposed by 
Castellani {\it et al.}\cite{Castellani}, where one electron-spin in the 
$a_1$ orbital couples to that of the other V ion of the pair
forming a singlet spin state with the neighboring site 
and $S=1/2$  spin degree of freedom 
arising from the other electron in the the $e^\pi$ orbital
remains in each V ion site.
The difference in the spin state of the V ions is essentially 
caused by their choice of the on-site exchange 
interaction value $\tilde{J}\sim 0.2$~eV, which is less than one third of our value.
For $D_{\rm trg}<0.09$~eV, where  energy of the $a_1$ orbital is 
lower than  that of the $e^\pi$ orbital,
one of electrons is in the
 $e^\pi$ orbital and the other is in the $a_1$ orbital in both V ion sites  
 (denoted by $e^\pi a_1;e^\pi a_1$).
On the other hand, for $D_{\rm trg}>0.34$~eV, two $e^\pi$ orbital are occupied in both V ion site 
(denoted by $e^\pi e^\pi;e^\pi e^\pi$).
 In between these two region, the ground state is
 a superposition of the $e^\pi e^\pi;e^\pi  a_1$ and $e^\pi  a_1;e^\pi e^\pi$
 states. 

For $0.09<D_{\rm trg}<0.34$, the ground state is doublet state and
their wave functions $|g\pm\rangle$ in large $U$ and 10$Dq$ limit 
can be approximately described  by 
\begin{eqnarray*}
           |g+\rangle&=&\frac{1}{\sqrt{2}}( 
               |e^\pi_{+\uparrow} {a_1}_\uparrow;e^\pi_{+\uparrow} e^\pi_{-\uparrow}\rangle
             +|e^\pi_{+\uparrow} e^\pi_{-\uparrow};e^\pi_{+\uparrow}{a_1}_\uparrow\rangle 
                           ), \\          
           |g-\rangle&=&\frac{-1}{\sqrt{2}}(
        |e^\pi_{-\downarrow} {a_1}_{\downarrow}; e^\pi_{+\downarrow} e^\pi_{-\downarrow}
            \rangle
      +|e^\pi_{+\downarrow} e^\pi_{-\downarrow}; e^\pi_{-\downarrow} {a_1}_\downarrow
            \rangle
                  ),\label{WavT}
\end{eqnarray*}
where $\uparrow$ ($\downarrow$) denotes the magnetic quantum number of the electron spin $s_z=1/2$ 
($s_z=-1/2$) and $|e_\pm^\pi\rangle =\frac{1}{\sqrt{2}}(\mp|e_u^\pi\rangle -{\rm i}|e_v^\pi\rangle)$ with $\langle e_\pm^\pi | L_z |e_\pm^\pi\rangle=\mp 1$.
These states have the  
orbital and spin angular momenta along the $z$-axis and the values per V ion are
$\langle g\pm|L_z|g\pm\rangle=\mp \frac{1}{2}$ and 
$\langle g\pm|S_z|g\pm\rangle =\pm 1$.
Since this degeneracy of ground state is magnetic,
there is no orbital degree of freedom.
This is very different result to the similar V ion pair model propose by Mila {\it et al.}\cite{Mila,Shiina}, 
where the orbital angular momentum is assumed to be small and thus the orbital degree of freedom
still remains apart from the spin degree of freedom. The ground state wave functions
of their model are written as
\begin{eqnarray*}
         |g \gamma;S^{\rm T}_z\rangle=\frac{1}{\sqrt{2}}( 
                     |e_\gamma^\pi a_1; e_u e_v^\pi \rangle 
                 + |e_u^\pi e_v^\pi ;e_\gamma^\pi a_1 \rangle
          )\otimes |S^{\rm T}=2;S^{\rm T}_z\rangle,\label{WavMila}
\end{eqnarray*}
where   $\gamma$ specifies the two kinds of the $e^\pi$ orbitals $u$ and $v$
and $|S^{\rm T}=2;S^{\rm T}_z\rangle$ is the total-spin wavefunction,
in which  $S=1$ Hund's spins of the two V ions are coupled ferromagnetically.
The ground state wavefunctions of our model $|g \pm\rangle$ can be describe
 in terms of the wavefunctions  $ |g \gamma;S^{\rm T}_z\rangle$ as
\begin{equation}
    |g\pm\rangle=\frac{1}{\sqrt{2}}(-{\rm i}|gu;S^{\rm T}_z=\pm 2\rangle\pm|gv;S^{\rm T}_z=\pm 2\rangle).
    \label{WavG}
\end{equation}
The wavefunctions of Mila's ground state
other than above two are lifted higher energies  ranging up to $\sim 0.027$~eV.

It has been reported from the x-ray diffraction
experiment that the distortion from the corundum structure to
the monoclinic structure at the PM $\to$ AFI transition involves abrupt expansion of 
the V-V bond length of the V ion pair and 
rotation of the V-V bond direction about 1.8$^\circ$ from the $\mib{c}_h$-axis
(within the $y$-$z$ plane in Fig.~\ref{tilting})\cite{Dernier}.
To take into account the effects of the lattice distortion, 
 the values of $dd\sigma$ and $dd\pi$ are reduced from those for
 the PM phase using the V-V distance $R$  dependence
 $(ddm) \propto 1/R^5$ proposed by Harrison\cite{Harrison} 
 and  the low symmetry crystal field are considered as in Table II.

In Fig.~\ref{a1occup}(b), the trigonal field $D_{\rm trg}$ dependence of 
the $a_1$ orbital occupation number in the ground state
for the AFI phase is shown.
The most different part compared to Fig.~\ref{a1occup}(a) is that
the occupation number decreases continuously as $D_{\rm trg}$ increases around 
$D_{\rm trg}\sim 0.3$~eV.
This indicates that the boundary between the $e^\pi e^\pi;e^\pi e^\pi$ and   
 $e^\pi a_1;e^\pi e^\pi$ ($e^\pi e^\pi;e^\pi a_1$)  states 
 is obscure and  around $D_{\rm trg}\sim 0.3$~eV, there is  
 $e^\pi e^\pi;e^\pi e^\pi$ and $e^\pi a_1;e^\pi  e^\pi$ ($e^\pi e^\pi ;e^\pi a_1$) 
mixed state.
This can be understood as follows.
Because of the symmetry lowering of the pair
caused by the tilting of the V-V bond direction from the trigonal axis, the $a_1$ 
and  $e^\pi_v$  orbitals hybridized in each other.
This leads to hybridization of the  $e^\pi e^\pi;e^\pi e^\pi$ and 
 $e^\pi a_1;e^\pi e^\pi$ ($e^\pi e^\pi;e^\pi a_1$) states around $D_{\rm trg}\sim 0.3$~eV,
 where energies of these two states are almost degenerate.
It is also seen in Fig.~\ref{a1occup}(b), the boundary 
 is shifted to much lower $D_{\rm trg}$ than that for the PM phase.
This shift of the boundary is due to reduction 
in the V-V hopping integral values.
 
 If we assume $D_{\rm trg}\sim 0.3$~eV,
 the abrupt change in the $a_1$ orbital occupation number
 is expected at the metal-insulator (M-I) transition.
 Such a large difference in the $a_1$ orbital occupation number 
 between the metal and insulating phases was indeed inferred from
  the measurement of  the linear dichroism of the V 2$p$ X-ray absorption spectrum\cite{Park}.
 Both the $e^\pi a_1$ to $e^\pi e^\pi$ ratio estimated from the measurement 
 for the AFI phase $e^\pi a_1$ : $e^\pi e^\pi=1:2$ and the PM 
 phase $e^\pi a_1$ : $e^\pi e^\pi=1:1$ are well reproduced with $D_{\rm trg}=0.28$~eV.
Since large change in the electronic state
 are caused by the lattice distortion, i.e., elongation and rotation of the V-V bond of the
 V ion pair, in this model, it is  expected that the lattice distortion is
 strongly related to the metal-insulator transition.
 It is also important to mention that there is also large
 difference in the weight of the $d^3d^1$ and $d^1d^3$ 
 configurations  between the ground states obtained for the two phases.
While  the weight is $w(d^3d^1)=w(d^1d^3)=3.5$\% for the AFI phase,
the weight for the PM phase is 6.9\%. Such large increase of the charge fluctuation
from the AFI phase to the PM phase in this model 
again indicates strong connection between
the monoclinic distortion and this metal-insulator transition.
\begin{figure}
\begin{center}
%\figureheight{3cm}
\epsfig{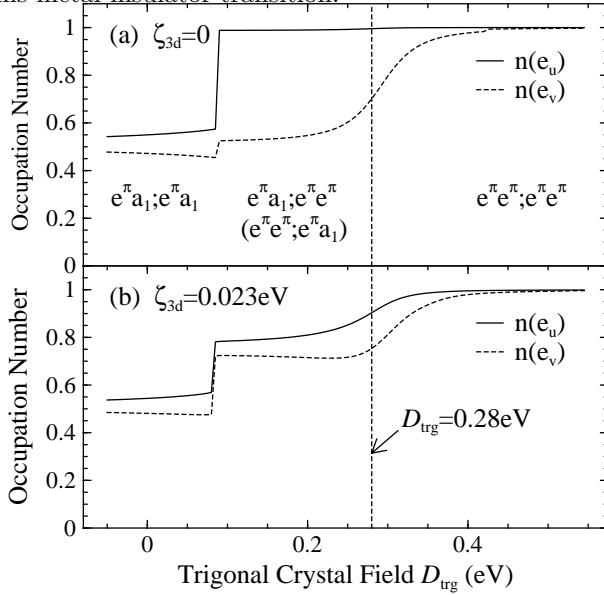}
\end{center}\par
\caption{
Trigonal crystal field $D_{\rm trg}$ dependence of 
the  $e_u$ ($e_v$) orbital occupation number $n(e_u)$ ($n(e_v)$) per V ion 
in the ground state without (a) and with (b) the 3$d$ spin-orbit interaction
for the AFI phase are shown.
The vertical line is positioned at $D_{\rm trg}=0.28$~eV, where the ground state
gives the ratio $e^\pi a_1:e^\pi e^\pi=1:2$ deduced from the XAS experiments for the AFI phase.
}\label{eoccup}
\end{figure}

To see the effects of the 3$d$ spin-orbit interaction
 on the ground state wavefunction in the AFI phase much clearly,
in Fig.~\ref{eoccup}, $D_{\rm trg}$ dependence of 
the  $e_u$  and $e_v$ orbital occupation numbers $n(e_u)$ and $n(e_v)$  per V ion
 without (a) and with (b) the 3$d$ spin-orbit interaction
are shown for the AFI phase. 
There is large difference between the two cases.
For $0.09<D_{\rm trg}<0.25$, without the spin-orbit interaction,
 the occupation numbers are $n(e_u)\sim 1$ and $n(e_v)\sim 1/2$
 and this shows that the ground state is mainly dominated by 
the $e^\pi_ua_1;e^\pi_ue^\pi_v$ ($e^\pi_ue^\pi_v;e^\pi_ua_1$)  configuration.
This also indicates that the monoclinic distortion will induce
the ferro V$_2$ molecular orbital order in the AFI phase,
which is proposed by Mila {\it et al.}\cite{Mila,Shiina}
On the other hand, in the same parameter range, with the spin-orbit interaction,
the occupation numbers are $n(e_u)\sim n(e_v)\sim 3/4$ and 
both two kinds of the configurations $e^\pi_ua_1;e^\pi_ue^\pi_v$ and $e^\pi_va_1;e^\pi_ue^\pi_v$
 are almost in equal weight.
 As was discussed before, this is because the strong coupling between the
 spin and orbital degeneracy and there are no orbital
 degree of freedom which is independent
 to the spin degree of freedom in this ground state and
 thus we do not expect usual Jahn-Teller instability.
Around $D_{\rm trg}\sim 0.3$~eV, where the $e^\pi e^\pi;e^\pi e^\pi$ state is mixed with 
the $e^\pi a_1;e^\pi e^\pi$ ($e^\pi e^\pi;e^\pi a_1$) state, the situation is similar: 
difference between  $n(e_u)$ and $n(e_v)$ is smaller with the spin-orbit interaction
than that without the interaction.

The ground state property with the spin-orbit interaction above can be 
 understood more clearly with the aide of an effective Hamiltonian 
 within the limit of large $U$ and 10$Dq$.
In this limit, the ground state is well described within
 restricted basis functions $|g e\rangle=|e_u^\pi e_v^\pi;e_u^\pi e_v^\pi \rangle$ and
$|g \gamma\rangle=\frac{1}{\sqrt{2}}(|e_\gamma^\pi a_1;e_u^\pi e_v^\pi \rangle
+ |e_u^\pi e_v^\pi;e_\gamma^\pi a_1 \rangle)$~~($\gamma=u,v$).
For convenience, here, $L'=1$ pseudo orbital 
angular momentum operator $\mib{L'}$ is introduced, where
the eigen functions $L'_z|g~m\rangle=m|g~m\rangle$ of $L'_z$ are defined by
 $|g~m=0\rangle=-|g e\rangle$ and
$|g~m=\pm1\rangle=(-{\rm i}|gu\rangle\pm|gv\rangle)/\sqrt{2}$.
The effective Hamiltonian can be expressed using $\mib{L}'$ and the $S^{\rm T}=2$ total spin operator
$\mib{S}^{\rm T}$ of the V ion pair as
\begin{eqnarray}
H_{\rm eff}&=&-\lambda\big\{ S^{\rm T}_zL'_z+\sqrt{2}(S^{\rm T}_xL'_x+S^{\rm T}_yL'_y)\big\}+\Delta {L'}_z^2\nonumber\\
    &+&\sqrt{2}C_{\rm A}(e^\pi_v,a_1) (L'_zL'_y+L'_yL'_z),\label{Hsl}
    \end{eqnarray}
where  $\lambda=\zeta_{3d}/4$ and $\Delta=D_{\rm trg}-\{(dd\sigma)^2+(dd\pi/3)^2\}/(U'-J)$.
Note that there are relations  between the real total orbital angular momentum
operator $\mib{L}$ and $\mib{L'}$ within this restricted basis: $L_x=-\sqrt{2}L'_x$,
$L_y=-\sqrt{2}L'_y$ and $L_z=-L'_z$.
If $C_{\rm A}(e^\pi_v,a_1)=0$, the system has uniaxial anisotropy and 
when $\Delta <-2\lambda$, the $z$-axis is the 
magnetic easy axis and  the ground state is $|g \pm\rangle=|g m=\pm 1\rangle|S^{\rm T}_z=\pm 2\rangle$, which is identical to that in eq.~(\ref{WavG}).
On the other hand, for $\Delta >-2\lambda$, the $x$-$y$ plane is the 
magnetic easy plane.
With finite $C_{\rm A}(e^\pi_v,a_1)$, the term $(L'_zL'_y+L'_yL'_z)$ causes hybridization 
between $|ge\rangle$ and $|gu\rangle$ and this rotate the magnetic easy
axis from the $z$-axis toward the $y$-axis.  
The ground state can be regarded as magnetic doublet
state having their magnetic moments along this easy axis with opposite directions in each other,
although breaking of the rotational symmetry around this easy axis causes a  
small energy splitting $< 0.5$~meV between these lowest two eigen states.

\subsection{Magnetic anisotropy}
\begin{figure}
\begin{center}
%\figureheight{3cm}
\epsfig{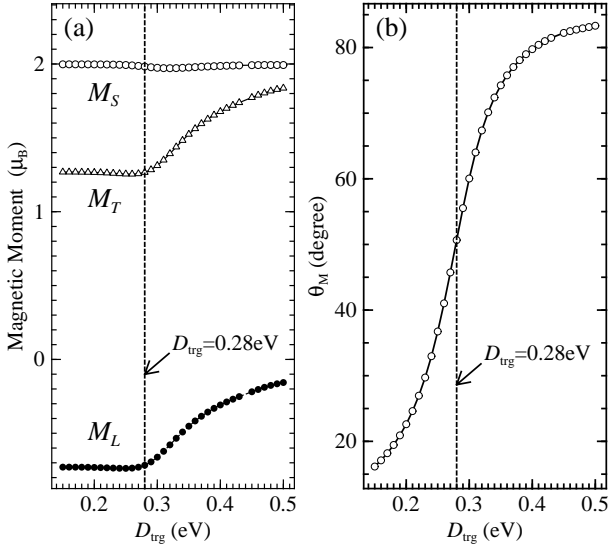}
\end{center}\par
\caption{
Sizes of the 
spin $M_{\rm S}$, orbital $M_{\rm L}$ and total $M_{\rm T}$ magnetic 
moments (a) and the canting angle 
of the total magnetic moment from the $z$-axis $\theta_{\rm M}$ (b)
 as a function of the trigonal crystal field $D_{\rm trg}$.
}\label{v2rotm}
\end{figure}

As was discussed in the previous subsection the admixture of the 
$e^\pi e^\pi;e^\pi e^\pi$ state in the ground state
lead to rotation of the magnetic easy axis.
To study such magnetic properties of the ground state, 
 a molecular field $H_{\rm M}$ with a canting angle $\theta_{\rm M}$
 with respect to the $z$-axis in the $y$-$z$ plane was further applied in the V ion pair model:
 \begin{eqnarray*}
  H_{\rm mol}
     =  H_{\rm M}\sum_\gamma \bigl\{&\cos \theta_{\rm M}& 
         (a^\dagger_{A\gamma\uparrow} a^{~}_{A\gamma\uparrow}
         -a^\dagger_{A\gamma\downarrow} a^{~}_{A\gamma\downarrow})\\
     -&{\rm i}\sin \theta_{\rm M}&
         (a^\dagger_{A\gamma\uparrow} a^{~}_{A\gamma\downarrow}
         -a^\dagger_{A\gamma\downarrow} a^{~}_{A\gamma\uparrow})\bigr\}.
\end{eqnarray*}
In the calculation, the molecular field with the magnitude $H_{\rm M}=0.01$~eV was
adopted and $\theta_{\rm M}$ was chosen
so as to give the lowest ground state energy. The magnetic moment is, therefore, parallel to
both the magnetic easy axis and the molecular field.
Note that the values and direction of the spin and orbital magnetic moments 
obtained here are not sensitive to the magnitude of the molecular field.

In Fig.~\ref{v2rotm}, the sizes of the 
spin $M_{\rm S}$, orbital $M_{\rm L}$ and total $M_{\rm T}$ magnetic 
moments per V ion (a) and $\theta_{\rm M}$ (b) are shown
as a function of $D_{\rm trg}$.
Both the directions of the magnetic moment and the V ion pair are
in the $y$-$z$ plane.
For the $e^\pi a_1;e^\pi e^\pi $ ($e^\pi e^\pi ;e^\pi a_1$)  ground state, the
direction of the magnetic moment is parallel to the $z$-axis (the corundum $\mib{c}_h$-axis).
However, as  the weight of the $e^\pi e^\pi ;e^\pi e^\pi$ configuration in the ground state
increases,  the  magnetic moment is canted
toward the $y$-axis
and finally reaches to the $y$-axis with a pure $e^\pi e^\pi;e^\pi e^\pi$ ground state.
Note that the V-V bond of the V ion pair is tilted toward opposite direction to the
magnetic moment, i.e., toward the negative direction
of the $y$-axis.
The large orbital magnetic moment $M_{\rm L} \sim 0.7\mu_{\rm B}$
is present in the $e^\pi a_1;e^\pi e^\pi $ ($e^\pi e^\pi ;e^\pi a_1$) ground state.
The orbital magnetic moment is about 40\% larger than  that
expected from the ground state wavefunction (see eq.~(\ref{WavT}) ) in the limit of
large $U$ and $10Dq$. 
 This deviation is mainly due to the
 presence of the small weight of configurations with occupied $e^\sigma$ orbitals
 arising from the 3$d$-3$d$ multipole interaction.
While $M_S$
 remains almost unchanged $M_S\sim 2 \mu_{\rm B}$ as the weight of  the $e^\pi e^\pi;e^\pi e^\pi$ configuration increases, 
$M_L$ decreases rapidly.
At $D_{\rm trg}=0.28$~eV, $M_{\rm T}\sim 1.2\mu_{\rm B}$ and $\theta_{\rm M}=51^\circ$ 
and these well agree with  the values obtained  in the neutron diffraction experiments
$M_{\rm T}\sim 1.2\mu_{\rm B}$ and $\theta_{\rm M}\sim 71^\circ$\cite{Moon}.
Strictly speaking, the magnetic moments of the two V ions are slightly tipped out 
from the $y$-$z$ plane toward opposite direction in each other. Although the x-component of the magnetic moments is an order of $\sim$ 0.01$\mu_{\rm B}$,  
the fact is important for the interpretation of the RXS measurements as will be discussed in \S~\ref{ANA111}.

\section{Electron-Lattice Interaction}

\begin{figure}
\begin{center}
%\figureheight{3cm}
\epsfig{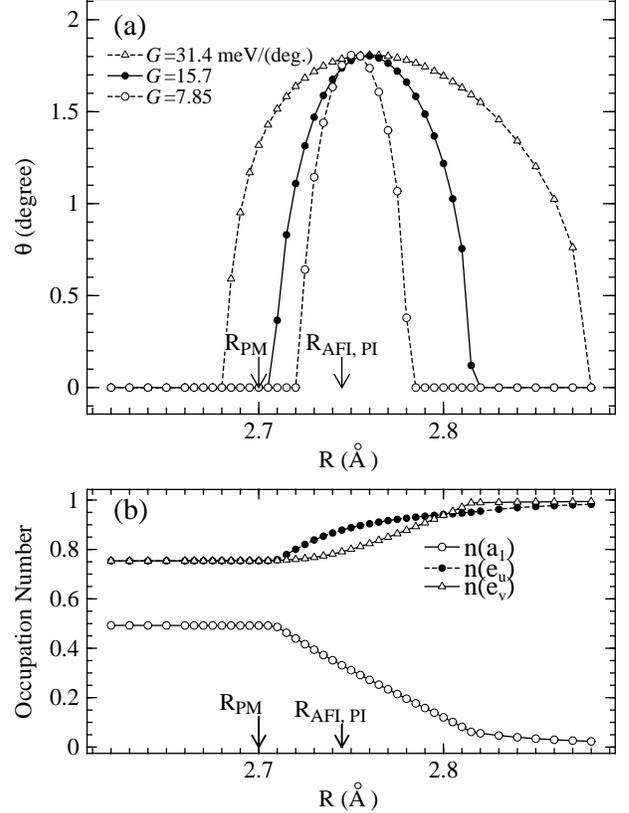}
\end{center}\par
\caption{In (a), tilting angle from the $\mib{c}_h$-axis as a function of the 
V-V distance of the V ion pair $R$ is shown for $G=7.85$, $15.7$ and $31.4$~meV/$(^\circ)$.
$R_{\rm AFI, PI}$ denotes the experimentally observed V-V distance of the V ion pair in both
the AFI and PI phases whereas $R_{\rm PM}$ shows that in
the PM phase.  In (b), the $a_1$, $e_u$ and $e_v$ orbital occupation numbers
calculated with $G=15.7$~meV/$(^\circ)$ are shown.
}\label{v2rotq}
\end{figure}
As was discussed in \S~\ref{GS}, 
the pure $e^\pi a_1;e^\pi e^\pi$ ($e^\pi e^\pi;e^\pi a_1$) ground state has only spin degree of freedom and
no orbital degree of freedom is left behind.
One would not, therefore, expect usual Jahn-Teller lattice instability in this system.
However, in the AFI phase, V$_2$O$_3$ exhibits the monoclinic lattice
distortion and this indicates that there must be some other kind of
lattice instability in this system.
Here, it will be shown that such instability indeed
appears when the energy of the $e^\pi a_1;e^\pi e^\pi$ ($e^\pi e^\pi ;e^\pi a_1$) and $e^\pi e^\pi;e^\pi e^\pi$ states
are almost degenerate.

In order to discuss interaction between the lattice deformation and the electronic state,
here, we consider one V ion pair in the crystal and in addition to the energy of 
the electronic state discussed in \S~\ref{model},
 the local lattice deformation energy caused by tilting of the V ion pair is included.
The energy of the V ion pair 
as a function of the
polar $\theta$ and azimuthal $\varphi$ angles of the tilting direction
would be written in the form:
\begin{eqnarray}
   E_{\rm tot}(\theta,\varphi)&=&E_{\rm elec}(\theta,\varphi)
   +\frac{K}{2}\theta^2+F\theta^3\sin 3\varphi+\dots,
\end{eqnarray}
where the first term denotes the ground state energy of the
Hamiltonian described in \S~\ref{model} and the second and third 
terms are the local lattice distortion energies.

Instead of including $F\theta^3\sin 3\varphi$ and much higher order 
trigonal potential terms of the lattice deformation,
for simplicity, we consider the lowest term
$\frac{K}{2}\theta^2$ only and 
 assume that tilting occurs only to the restrict
directions of $\varphi=-\frac{\pi}{2}$, $-\frac{\pi}{2}\pm \frac{2\pi}{3}$, where
displacements of each V ion of the pair is parallel to
one of the three V-V bond in the honeycomb layer.
As a result, the V-V bond along this direction is elongated and
those in the other two directions contract.
Note that this tilting direction is consistent with  that observed in the monoclinic
lattice distortion in the AFI phase.
In such lattice distortion, both the $e^\pi_u$ and $e^\pi_v$ orbitals
can be hybridized with the $a_1$ orbital and 
corresponding low symmetry crystal field parameters, which
are linear functions in $\theta$ for small lattice distortion, can be written in the form:
\begin{eqnarray}
C_{\rm A}(e_u^\pi,a_1)=\frac{G\theta}{\sqrt{2}}\cos\varphi,~~
C_{\rm A}(e_v^\pi,a_1)=\frac{G\theta}{\sqrt{2}}\sin\varphi.
\end{eqnarray}
In the calculation, $D_{\rm trg}=2.8$~eV was adopted 
and the V-V distance $R$ dependence for the $d$-$d$ hopping integrals
 $dd\sigma,dd\pi \propto 1/R^5$ was assumed.

In Fig.~\ref{v2rotq}, 
the tilting angle at the minimum of $E_{\rm tot}$
as a function of the V-V distance of the V ion pair 
$R$ is shown for three different crystal
fields $G=7.85$ (open circles), $15.7$ (closed circles) and $31.4$~meV/($^\circ$) (triangles).
 The coefficient $K$ was chosen so that maximum tilting angle is 1.8$^\circ$ and
 those for $G=7.85$, $15.7$ and $31.4$~meV/($^\circ$)
 are $K=2.60$, $6.50$
 and $15.7$~meV/$(^\circ)^2$, respectively. 
The V-V distances $R_{\rm AFI, PI}$ and $R_{\rm PM}$ 
are the experimentally observed values
in the insulating phases and in the PM phase, respectively\cite{Dernier}. 
 The rotation of the V ion pair occurs only when
the energy of the two kinds of the states $e^\pi a_1;e^\pi e^\pi$ 
($e^\pi e^\pi;e^\pi a_1$)  
and $e^\pi e^\pi;e^\pi e^\pi$
are nearly degenerate 
and takes the maximum value around $R=2.750\stackrel{\tiny\circ}{\rm A}$, 
where the two kinds of the states are just degenerate.
Note that the energy difference between the two 
can be given by $\sim D_{\rm trg}-\{(dd\sigma)^2+(dd\pi/3)^2\}/(U'-J)$.

This instability toward the lattice distortion can be explained by the similar mechanism
to the Jahn-Teller effects.
If we include the lattice distortion energy, the effective Hamiltonian eq.~(\ref{Hsl})
generalized for the V ion pair with an arbitrary tilting angle $(\theta,\varphi)$ can be
written as
\begin{eqnarray}
H_{\rm eff}&=&-\lambda\big\{S^{\rm T}_zL'_z+\sqrt{2}(S^{\rm T}_xL'_x+S^{\rm T}_yL'_y)\big\}+\Delta {L'}_z^2\nonumber\\
    &+&G\theta\{\cos\varphi (L'_zL'_x+L'_xL'_z)+\sin\varphi(L'_zL'_y+L'_yL'_z)\}\nonumber\\
    &+&\frac{K}{2}\theta^2+F\theta^3\sin 3\varphi+\dots. \label{Heff2}
\end{eqnarray}
When the energies of the these two kind of states $|ge\rangle$ and  
$|g\gamma\rangle$ ($\gamma=u,v$) are degenerate or nearly degenerate, the term linear in $\theta$,
which couples the states $|ge\rangle$ and $|g\gamma\rangle$, causes  
rotation of the V ion pair in the same way as the Jahn-Teller effects.
The condition for instability toward tilting
is approximately obtained for small $\lambda$ and $|F|$ as $|\Delta| < 2G^2/K$.

Since the energy gain from this electron-lattice interaction is maximum
at the V-V distance where the energies of the two kinds of the states are nearly degenerate,
if this mechanism is the cause of the monoclinic distortion in the AFI phase,
the V-V distance of the pair in this phase should be at this distance.
Indeed, the value $R=2.755\stackrel{\tiny\circ}{\rm A}$ giving the maximum tilting angle is  very near to
the value  $R_{\rm AFI,PI}=2.745\stackrel{\tiny\circ}{\rm A}$, where
 the V 2$p$ XAS experimental results on the $e^\pi e^\pi $ and $e^\pi a_1$ configuration 
 ratio for the AFI phase  $e^\pi a_1 : e^\pi e^\pi=1:2$ is reproduced\cite{Park}.
The monoclinic lattice distortion in the AFI phase thus can be understood
by this mechanism.
On the other hand, the V-V distance is much shorter in the PM phase
$R_{\rm PM}\sim 2.700\stackrel{\tiny\circ}{\rm A}$ and
if  $|G| <16$~meV, the lattice instability does not occur at this distance.
In this situation, pure $e^\pi a_1;e^\pi e^\pi$ ($e^\pi e^\pi;e^\pi a_1$) state is expected
in this phase. This also agrees with the V 2$p$ XAS experiment,
where $e^\pi a_1:e^\pi e^\pi=1:1$ is obtained in the PM phase.

Although the experimentally observed V-V distance is the same to that for the AFI phase,
in the PI phase, no monoclinic lattice distortion has been found.
This can be explain as follows.
In the PI phase, each V ion pair is also tilted from the 
$\mib{c}_h$-axis similar to the AFI phase.
However, their tilting directions are disordered and fluctuate among the three 
stable directions $\varphi=-\frac{\pi}{2}$, $-\frac{\pi}{2}\pm \frac{2\pi}{3}$, resulting in no monoclinic lattice distortion. 
This assumption is consistent with the V 2$p$ XAS experimental results,  where the
weight of the $e^\pi e^\pi$ configuration
is larger than that of the $e^\pi a_1$ configuration
both in the PI and  AFI phases in their initial state, 
indicating admixture of the $e^\pi e^\pi;e^\pi e^\pi$ configuration.
If this assumption is correct, the AFI $\to$ PI transition
is accompanied by order to disorder transition in the tilting directions of the V ion pairs.
As will be discussed in the next section, this ordering of the tilting direction in the AFI phase
can be explained by a model in which this local lattice distortion
and the magnetic exchange interaction between the V ion pairs are considered.

From above discussion, it is expected that 
there are at least two local minima
for the free energy of the real system as a function of the V-V distance of the pair.
One is positioned at $R_{\rm PM}\sim 2.700\stackrel{\tiny\circ}{\rm A}$, where the energy difference of the
two kinds of the state $e^\pi a_1;e^\pi e^\pi$ and $e^\pi e^\pi;e^\pi e^\pi$ 
is large and thus no electron-lattice coupling takes place,
and the other one is located at $R_{\rm AFI,PI}\sim 2.745\stackrel{\tiny\circ}{\rm A}$,
where energy of the two kinds of the states is degenerate
and Jahn-Teller like lattice distortion is 
the cause of the energy lowering at this V-V distance.
The expansion of the V-V distance in the AFI and PI phases as compared to that
in the PM phase can be understood as a consequence of the 
large change in the 3$d$ orbital occupation owing to the tilting of the V ion pair.
Since mixture of the $e^\pi e^\pi;e^\pi e^\pi$ state in the ground state reduces
electron occupation of the $a_1$ orbital, which point toward
the direction of the V-V bond, the nuclear charges of the V ions of the pair
is much less screened by the 3$d$ electron in the V-V bond direction, resulting in elongation
of the V-V bond.

Since both in the AFI and PI phases, the electron-lattice coupling strongly influences
the orbital occupation and charge fluctuation,
the PM $\to$ AFI and PM $\to$ PI transitions can not be regarded
as usual Mott transition.
In \S~\ref{GS}, it is found that when the lattice distortion occurs,
charge fluctuation between the two V ions is strongly suppressed.
The expansion of the V-V distance and resultant reduction of the 
$d$-$d$ hybridization strength due to this electron-lattice coupling in the 
insulating phases is probably the main cause of this 
metal-insulator transition.

\section{Magnetic Interaction between V Ion Pairs}

\begin{figure}
\begin{center}
%\figureheight{3cm}
\epsfig{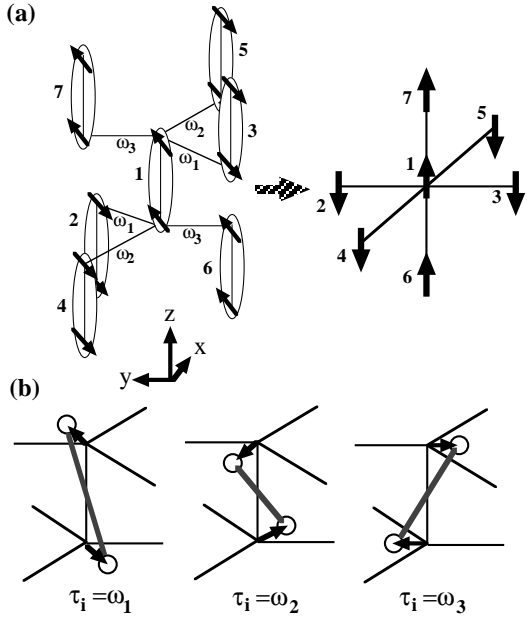}
\end{center}\par
\caption{
Mapping of V ion pairs in the corundum lattice on the simple cubic lattice (a)
and three possible directions corresponding to $\tau_i=\omega_1$, 
$\omega_2$ and $\omega_3$ of the tilting of the V ion pair from 
the $\mib{c}_{\rm h}$-axis (b) are
drawn. On  left hand side of Fig. (a), the ovals show the V ion pairs and the allows represent
the spin directions of the V ions and on right hand side,
the directions of the pseudo spin $\sigma_i=\pm 1$ for V ion pairs are indicated by the arrows.
In Fig. (b), the circles and arrows represent the V ions and their displacement due to
the tilting of the V ion pair, respectively.
}\label{spin}
\end{figure}
In the previous section, we have discussed about the 
local lattice distortion effects on the electronic states within 
individual V ion pairs.
Here, we will consider about 
the exchange interaction between the V ion pairs 
in its relation with the local lattice distortion
 in order to discuss the AFI $\to$ PI transition observed in V$_{1-x}$Cr$_x$O$_3$.
The transition is accompanied by the monoclinic to 
corundum lattice structure transition
and in the ultrasonic wave velocities measurements,
the $C_{44}$ elastic constant was estimated to soften almost at the transition
temperature, showing the nearly second order 
nature of the transition\cite{Yang,Yethiraj}. 
Contrary to this, in the AFI to PM transition 
 such softening of $C_{44}$ is not observed and the transition is of the first order.
In the magnetic thermal-neutron scattering experiments,
abrupt change in short range spin correlation function
at the transition temperature was observed.\cite{Bao,Bao2}

In the AFI phase, the monoclinic lattice distortion and the magnetism
are connected, forming unique AF magnetic order.
In the monoclinic distortion,
all the V ion pairs, which is parallel to the $\mib{c}_h$-axis 
in the corundum structure are tilted $\sim 1.8^\circ$ in the same direction\cite{Dernier}.
With this displacement of the V ions, 
the honeycomb lattice, which has three hold rotational symmetry in the corundum
structure, is distorted and V-V bonds in one of the three directions in the layer
are elongated and those in the other two directions contract (see Fig.~\ref{tilting}).
The magnetic moments between two V ions with the
elongated V-V bond on the honeycomb lattice 
are ferromagnetically coupled and those with the
contracting V-V bond are antiferromagnetically coupled.

In the theories proposed by Castellani {\it et al.}\cite{Castellani} and Mila {\it et al.}\cite{Mila,Shiina},
the orbital ordering is an essential ingredient for the formation 
of this unique AF magnetic order in V$_2$O$_3$, where exchange coupling constant between the 
spins of V ions depends on the orbital occupation of the V ions.
Rice has pointed out that the abrupt change in the spin correlations at the AFI $\to$ PI transition
and the short-range spin fluctuations in the PI phase can be explained in the presence of the
orbital degree of freedom\cite{Rice}.
Contrary to this, Ezhov {\it et al.}\cite{Ezhov} performed 
 LDA+$U$ band structure calculations and 
obtained a solution with the unique AF magnetic order 
 as the lowest total energy magnetic state, 
assuming the monoclinic lattice distortion.
Since the orbital occupation in each V ion is $e^\pi e^\pi$,
there are no orbital order in this AF magnetic state.
In the present theory, since strong coupling between the spin and orbital
degrees of freedom due to the spin-orbit interaction,
orbital order does not exist in the ground state.
Here, it is assumed that the coupling constant of the exchange interaction is 
a function of  the V-V bond length and not depends on
the orbital occupation of the V ions.

\subsection{Effective spin-lattice Hamiltonian}

The effective spin Hamiltonian considered here consists
of  the exchange interaction between the nearest neighbors of the 
V ion pairs 
and the local lattice distortion energy in each V ion pair.
Each V ion pair has six nearest neighbors and
as depicted in Fig.~\ref{spin}, the network of the V ion pairs 
can be mapped onto the simple cubic lattice. 
 Two spin directions of the doublet ground state
 on the i-th V ion pair 
is projected onto Ising like 
pseudo-spin with $\sigma_i=\pm 1$ at the corresponding 
lattice point $i$ of this simple cubic lattice.
The coupling constant of the exchange interaction is assumed to be a function of 
the displacement of the V-V bond length due to
tilting of the V ion pairs.
The effective spin Hamiltonian then can be written as
\[
H_{\rm S}=-J\sum_{\langle i,j\rangle}
\{(\theta_i \mib{\tau}_i + \theta_j \mib{\tau}_j)\cdot\mib{l}_{ij}-C\}\sigma_i\sigma_j
+\sum_i \bigl(\frac{K}{2}{\theta_i}^2-G\theta_i \bigr),
\]
where the first term represents the exchange interaction 
between the pseudo-spins of the V ion pairs and the second term describes
the local lattice distortion energy in each V ion pairs.
$\theta_i\ge 0$ represents tilting angle of the i-th V ion pair
and two dimensional unit vector $\mib{\tau}_i$ denotes
the direction of the tilting projected on the honeycomb layer.
 Here, we assume that each V ion pair is tilted
in such a direction that 
each V ion in both end is parallel to 
one of the three V-V bond in the honeycomb layer and elongates this V-V bond.
$\mib{\tau}_i$ is, therefore, in one of these three directions:
$\mib{\omega}_1=(\sqrt{3}/2,1/2)$, $\mib{\omega}_2=(-\sqrt{3}/2,1/2)$ or $\mib{\omega}_3=(0,-1)$.
 $\mib{l}_{ij}=\mib{\omega}_1$, $\mib{\omega}_2$ or $\mib{\omega}_3$ 
 indicates the direction of the V-V bond between the i- and j-th V ion pairs.
 Thus, the term $(\theta_i \mib{\tau}_i + \theta_j \mib{\tau}_j)\cdot\mib{l}_{ij}$
 is proportional to the elongation of the V-V distance
 between the i- and j-th V ion pairs.
 
The AF magnetic order observed in V$_2$O$_3$ corresponds to the C-type AF magnetic order
in this model, where the nearest neighbor
spins along one of the lattice line in the simple cubic lattice are
coupled ferromagnetically and those along the other two are antiferromagnetically
coupled (see Fig.~\ref{spin} (a)).
For this AF magnetic order, the energy per V ion pair is easily obtained as a function of a uniform
 distortion $\theta=\theta_i$
\[
   E(\theta)=-|J|C-(G+4|J|)\theta+\frac{K}{2}\theta^2
\]
and the energy has minimum at $\theta_{\rm min}=\frac{G+4|J|}{K}$
\[
     E(\theta_{\rm min})=-|J|C-\frac{1}{2}\frac{(G+4|J|)^2}{K}.
\]

For this AF magnetic state being stable,  the exchange coupling constant 
$J(2\theta_{\rm min}-C)$ for  the long V-V length should be positive
and that for the short V-V length $-J(\theta_{\rm min}+C)$ should be negative. 
Thus at least the conditions $J<0$ and  $2\theta_{\rm min}>C>-\theta_{\rm min}$
must be satisfied for this AF magnetic state being the ground state.

\subsection{AFI to PI phase transition}
\begin{figure}
\begin{center}
%\figureheight{3cm}
\epsfig{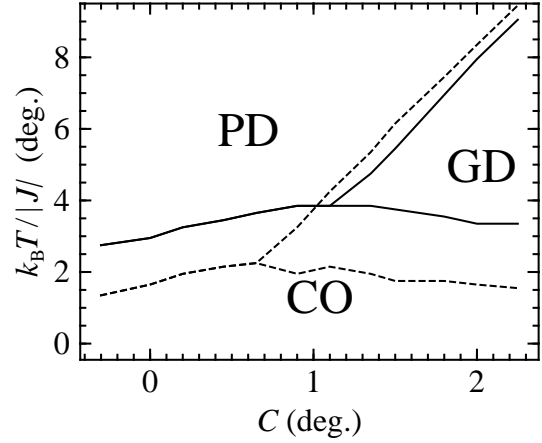}
\end{center}\par
\caption{
Phase diagram of the effective spin Hamiltonian for $G/|J|=3$ (solid line)
and $G/|J|=0$ (dashed line).
CO denotes the C-type AF magnetic order with ordered tilting direction of V ion pair.
GD and PD are the G-type AF magnetic order and paramagnetic order
with disordered tilting direction, respectively.
}\label{diagram}
\end{figure}

To study finite temperature properties of this effective Hamiltonian,
Monte Carlo simulations were performed with a $24\times 24\times 24$ periodic lattice 
for various values of $C$.
Figure \ref{diagram} shows the phase diagram as a function of $C$ and $k_{\rm B}T/|J|$
obtained from the calculation.
The adopted value of $G$ is $G=15.7$~meV/($^\circ$) and from the experimental
facts $\theta_{\rm min}\sim 1.8^\circ$ and $T_{\rm N}\sim 155$~K,
the other parameter values are estimated to be $J\sim -4.0$~meV /($^\circ$),
$K/|J|\sim$ 3.9 $(^\circ)^{-1}$ and $G/|J|\sim 3.0$.
The solid and dashed lines represent the phase boundaries calculated with ($G/|J|=3$)
and without ($G/|J|=0$) the electron-lattice coupling, respectively.
In the range of $-0.3^\circ<C<2.25^\circ$ as shown in the figure, the low temperature phase is that
with the C-type AF magnetic order and the V ion pairs are tilted and ordered 
in the same direction in this phase (denoted by CO). This phase
corresponds to the AFI phase in the real material.
The high temperature  phase is paramagnetic and
the tilting direction of the V ion pairs are disordered (denoted by PD) and
this phase corresponds to the PI phase in the real material.
For $C>1.1^\circ$, there is an intermediate temperature phase with
 the G-type AF magnetic order,
 where all the nearest neighbor spins are antiferromagnetically coupled, 
 with disordered tilting directions of the V ion pairs (denoted by GD).
The transition from the CO phase to the PD phase
is of the second order whereas  the CO $\to$ GD and GD $\to$ PD
phase transitions are of the first order.
It is also seen in the figure that the CO phase is further stabilized with the 
electron-lattice interaction (solid line) than without it (dashed line), 
showing importance of the
interaction for the formation of the C-type AF magnetic order.
\begin{figure}
\begin{center}
%\figureheight{3cm}
\epsfig{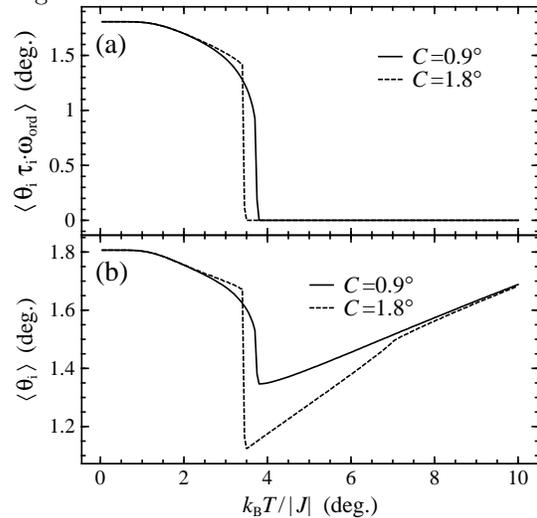}
\end{center}\par
\caption{
Order parameter for the tilting of the V ion pairs 
$\langle \theta_i \mib{\tau}_i\cdot \mib{\omega}_{\rm ord} \rangle$ and  the tilting angle $\langle \theta_i \rangle$ as a function of 
temperature calculated with $C=0.9^\circ$ (solid line) and $C=1.8^\circ$ (dashed line).
}\label{tau}
\end{figure}

\begin{figure}
\begin{center}
%\figureheight{3cm}
\epsfig{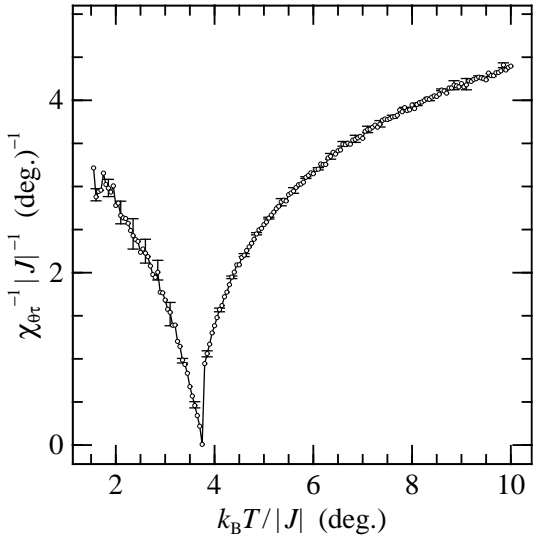}
\end{center}\par
\caption{
$1/\chi_{\theta\tau}(T)$ is plotted with error bars
 for $C=0.9^\circ$ as a function of temperature.
}\label{invxi}
\end{figure}

In Fig.~\ref{tau}, the order parameter of the tilting of V ion pairs 
$\langle \theta_i \mib{\tau}_i\cdot \mib{\omega}_{\rm ord} \rangle$ and  the tilting angle $\langle \theta_i \rangle$ as a function of 
temperature are shown for $C=0.9^\circ$ (solid line) and $C=1.8^\circ$ (dashed line), where $\mib{\omega}_{\rm ord}$
is one of the three possible tilting directions $\mib{\omega}_l$ ($l=1,2,3$)
in which the V ion pairs are ordered.
It is seen in the figure that the CO $\to$ PD transition is of the 
second order and the CO $\to$ GD transition is
of the first order. The tilting angle $\langle \theta_i \rangle$ in the PD
phase is much larger than that in the GD phase, showing stronger dynamic lattice
distortion in the PD phase.
The order parameter of the tilting
$\langle \theta_i\mib{\tau}_i\cdot\mib{\omega}_{\rm ord}\rangle$ 
is expected to be approximately proportional to the degree 
of the monoclinic  lattice distortion corresponding to the strain $e_4$ (see Fig.~\ref{tilting}).
The reciprocal of the quantity defined by 
$\chi_{\theta\tau}(T)=(\langle \theta_i\mib{\tau}_i\cdot\mib{\omega}_{\rm ord}\rangle^2
-\langle \bigl(\theta_i\mib{\tau}_i\cdot\mib{\omega}_{\rm ord}\bigr)^2\rangle)/k_{\rm B}T$
is therefore proportional to the elastic stiffness constant $C_{44}$.
In Fig.~\ref{invxi}(c), $1/\chi_{\theta\tau}(T)$ is plotted as a function of temperature.
Both above and below $T_{\rm N}$,
$1/\chi_{\theta\tau}$  decreases as temperature approaches to $T_{\rm N}$.
Since the transition is of the second order,
the value of $1/\chi_{\theta\tau}$ is zero just at $T_{\rm N}$.
This temperature dependence of  $1/\chi_{\theta\tau}$ is consistent with the 
 behavior of $C_{44}$ observed in ultrasonic wave velocity measurements\cite{Yang} 
 and also in neutron scattering experiments\cite{Yethiraj}
 on Cr doped V$_2$O$_3$, where 
$C_{44}$ softens with decreasing temperature in the PI phase
and approaches zero near the PI $\to$ AFI transition temperature.

\begin{table}
\caption{Spatial correlation functions $\langle \sigma_i \sigma_j \rangle$
and $\langle \theta_i  \mib{\tau}_i\cdot\theta_j\mib{\tau}_j \rangle$
between the first, second  and third nearest neighbors
along a lattice line are shown above and below the CO $\to$ PD and  CO $\to$ GD
phase transitions.
In the CO phase, those along two different lattice lines, parallel and
perpendicular to the ferromagnetic plane of the C-type AF order, are
listed.
\label{correlation}}
\begin{tabular}{rcccc}\hline
\multicolumn{5}{l}{$C=0.9^\circ,~K/|J|=3.9~(^\circ)^{-1},~G/|J|=3$}\\
\multicolumn{5}{l}{~~$k_{\rm B}T/|J|=3.65^\circ$ (CO phase)} \\
    &\multicolumn{2}{c}{$\langle \sigma_i\sigma_j \rangle$}&
    \multicolumn{2}{c}{$\langle \theta_i \mib{\tau}_i \cdot\theta_j\mib{\tau}_j\rangle$}  \\ \hline
1st &0.678 &$-0.735$&  1.21 &1.16 \\
2nd &0.670 & ~~$0.702$&1.07& 1.09 \\
3rd & 0.681 &$-0.691$&  1.03 &1.05\\ 
\hline
\multicolumn{5}{l}{~~$k_{\rm B}T/|J|=3.9^\circ$  (PD phase)} \\ 
    &\multicolumn{2}{c}{$\langle \sigma_i\sigma_j \rangle$}&
    \multicolumn{2}{c}{$\langle \theta_i\mib{\tau}_i \cdot\theta_j \mib{\tau}_j\rangle$}  \\     \hline
1st &\multicolumn{2}{c}{ $-0.191$}&\multicolumn{2}{c}{  0.170}  \\
2nd &\multicolumn{2}{c}{ ~~$0.141$}& \multicolumn{2}{c}{0.032 } \\
3rd &\multicolumn{2}{c}{ $-0.042$}& \multicolumn{2}{c}{0.005} \\ \hline
\multicolumn{5}{l}{$C=1.8^\circ,~K/|J|=3.9~(^\circ)^{-1},~G/|J|=3$}\\
\multicolumn{5}{l}{~~$k_{\rm B}T/|J|=3.3^\circ$ (CO phase)} \\ 
    &\multicolumn{2}{c}{$\langle \sigma_i\sigma_j \rangle$}&
    \multicolumn{2}{c}{$\langle \theta_i \mib{\tau}_i \cdot\theta_j\mib{\tau}_j\rangle$}  \\ \hline
1st &0.964 &-0.969&  2.02 &2.12 \\
2nd &0.965 & ~~0.966&2.10& 2.04 \\
3rd & 0.964 &-0.965&  2.07 &2.04\\ 
\hline
\multicolumn{5}{l}{~~$k_{\rm B}T/|J|=3.8^\circ$  (GD phase)} \\ 
    &\multicolumn{2}{c}{$\langle \sigma_i\sigma_j \rangle$}&
    \multicolumn{2}{c}{$\langle \theta_i \mib{\tau}_i \cdot\theta_j\mib{\tau}_j\rangle$}  \\ \hline
1st &\multicolumn{2}{c}{ $-0.957$}&\multicolumn{2}{c}{ $-0.001$}  \\
2nd &\multicolumn{2}{c}{~~$0.955$}&\multicolumn{2}{c}{~~$0.007$} \\
3rd &\multicolumn{2}{c}{ $-0.955$}&\multicolumn{2}{c}{$-0.002$} \\ \hline
\end{tabular}
\end{table}
In Table \ref{correlation}, 
spatial correlation functions $\langle \sigma_i \sigma_j \rangle$
and $\langle \theta_i  \mib{\tau}_i\cdot\theta_j\mib{\tau}_j \rangle$
between the first, second  and third nearest neighbors
along a lattice line are listed.
Those for temperatures above and below the CO $\to$ PD transition
calculated with $C=0.9^\circ$ are shown and for comparison, 
those for the CO $\to$ GD transition calculated with $C=1.8^\circ$ are also listed.
In the CO phase, those along two different lattice lines, parallel and
perpendicular to the ferromagnetic plane of the C-type AF order, are
shown.
In the PD phase, there is short range G-type AF spin correlation, which
 is consistent with recent neutron scattering experimental results\cite{Bao,Bao2}.
In addition to this, short range order of tilting directions is also present in this phase.
This sudden change in the spin correlation, which is unusual for order-disorder magnetic 
transition can be explained as follows.
In the PD phase, 
spins and the local lattice deformation are still dynamically coupled 
and short range C-type AF order with local lattice distortion survives.
However, since the short range C-type AF order with
three different directions of the ferromagnetic plane are coexistence,
 we find, in average, the G-type AF spin correlation as seen in Table III.
In this way, the abrupt change in the spin correlation functions
observed at the AFI $\to$ PI transition 
in the neutron scattering experiments is explained as a result of the
 simultaneous order to disorder 
transition of both the tilting directions and spins of the V ion pairs.
On the other hand, in the GD phase, 
since lattice distortion is not required for the stabilization of the G-type
AF spin order, the correlation functions between the tilting directions $\langle \theta_i  \mib{\tau}_i\cdot\theta_j\mib{\tau}_j \rangle$ is almost zero.
This is in contrast to the G-type AF spin correlation in the PD phase
originating from the short range C-type AF order with the dynamic spin-lattice coupling.

\section{Resonant X-ray Scattering Analysis}
Resonant x-ray scattering (RXS) has received much attention recently,
as a technique for extracting direct information on magnetic\cite{Gibbs,Hannon,McWhan3} and orbital\cite{Hirota} order.
Possibility of observing the orbital ordering in V$_2$O$_3$ by RXS 
have been pointed out by Fabrizio {\it et al.}\cite{Fabrizio} They made predictions on  
the light polarization and the azimuthal angle of the scattering plane dependence of 
the resonant Bragg reflection intensities,
 assuming the orbital order proposed by Castellani {\it et al.}\cite{Castellani}
The RXS measurements on V$_2$O$_3$\cite{Paolasini} 
and its Cr doped alloys\cite{Paolasini2} have been performed by Paolasini {\it et al.}
They insisted that in the AFI phase, there are resonant Bragg reflection peaks arising from 
the orbital ordering in addition to those caused by the AF magnetic ordering.
On the other hand, Lovesey and Knight claimed that all these Bragg peaks can be attributed
to the AF magnetic ordering and are associated with the orbital magnetic moment
and the octupole moment of a V ion\cite{Lovesey}.
In this section, discussion will be mainly focused on these resonant Bragg peaks, especially
(111) reflection peak, whose origin is still controversial.
It will be shown that the light polarization and the azimuthal angle
dependence of the (111) Bragg reflection intensity can be explain within the V ion pair model 
and is attributed to not the orbital ordering but the antiferro (AF) magnetic ordering.

\subsection{$1s \rightarrow 4p$ dipole X-ray scattering process}
\begin{figure}
\begin{center}
%\figureheight{3cm}
\epsfig{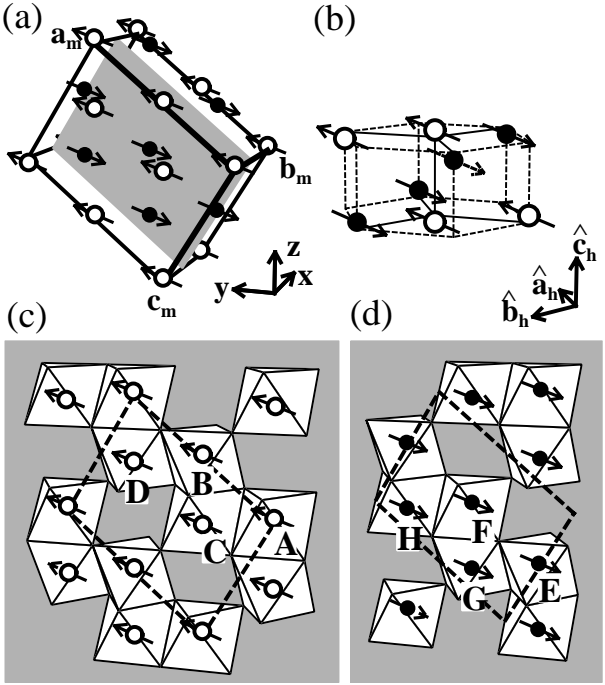}
\end{center}\par
\caption{Antiferromagnetic and crystal structures of V$_2$O$_3$
in the AFI phase.
The white and black circles denote V ions and the arrows indicate 
direction of their magnetic moments.
In (a),  the monoclinic unit cell and the magnetic structure are shown.
The  magnetic moments are ferrromagnetically coupled in each monoclinic (010)
layer, altering their direction between adjacent layers.
The V ion pair and neighboring six V ions are depicted in (b).
In (c) and (d),  V ions belonging to two different $(010)$ layers 
 are separately depicted with surrounding oxygen octahedra. 
  \label{crymag}
}
\end{figure}

In this subsection, we will discuss the $1s \rightarrow 4p$ dipole process, which is dominant 
 process in the RXS at the V $K$-edge absorption region.
The x-ray scattering amplitude $f_{\rm DD}(\omega)$ is expressed as
\begin{equation}
 f_{\rm DD}(\omega)= \sum_l e^{ {\rm i}(\mib{q}-\mib{q'})\cdot\mib{R}_l }
     ( { \mib{\varepsilon}' }^* \cdot F^l(\omega) \mib{\varepsilon}) \label{fomega},
\end{equation}
where $\mib{q}$ and $\mib{\varepsilon}$ are the wave vector
and the polarization vector for the
the incoming photon, $\mib{q'}$ and $\mib{\varepsilon'}$ are
those for the outgoing photon and $\mib{R}_l$ denotes position of the $l$-th V ion site.

Here, the $3\times 3$ tensor $F^l(\omega)$ is defined by
\begin{equation}
 F^l(\omega)=q^2\langle{\rm  g}| \mib{P} \frac{1}{E_g+\omega-{H}_{\rm m} 
                             +{\rm i}\Gamma_{\rm m}/2}
                  {^t\!\mib{P} }|{\rm g} \rangle,\label{tensor}
\end{equation}
where $|g\rangle$ and $E_g$ are the ground state wave functions and its energy, 
$H_{\rm m}$ and $\Gamma_{\rm m}$ are the Hamiltonian and a life-time term for 
the intermediate state reached by the $1s \to 4p$ transition 
 and $\mib{P}=e\mib{r}$ is the electric dipole moment operator.
Since delocalized feature of the 4$p$ orbital, the intermediate-state-multiplet dependence
of RXS is not expected to be important.
Thus, here we will consider the integrated value of $F^l(\omega)$
over the photon energy $\omega$:
\begin{equation}
    F^l_{i,j}=\int F^l_{i,j}(\omega)\,d\omega.
\end{equation}
Note that $F^l_{x,x}$, $F^l_{y,y}$ and $F^l_{z,z}$ correspond to 
the $1s \rightarrow 4p$ dipole transition probabilities integrated by $\omega$ at site $l$
with x-, y- and z- linearly polarized light, respectively.

Before discussing rather complicate RXS in V$_2$O$_3$,
first of all, let us consider the most simplest  antiferromagnet, where
 the unit cell with one magnetic ion site in the absence of magnetic order 
 is doubled by the presence of the antiferromagnetic order and 
 two magnetic sublattices A and B with opposite direction of the magnetic moment 
 in each other are assumed.
In this case,  $F^l$ for an ion site belongs to the sublattice A
(here we denote it by $F^{\rm A}$)
can be obtained from that belongs to the sublattice B (denoted by $F^{\rm B}$)
by replacing the ground state wavefunction $|g \rangle$ in eq.~(\ref{tensor})
with $\Theta T|g \rangle$, where $\Theta$ represents the time reversal operator
and $T$ denotes the translation operator, which transforms ion sites
 belong to the sublattice A into those belong to the sublattice B.
This immediately leads to the relation:
\[
F^{\rm A}=F^{{\rm B} *}.
\]
For resonant Bragg reflection corresponding to 
the antiferromagnetic order, the integrated scattering amplitude $f_{\rm DD}$ is
\begin{eqnarray}
      f_{\rm DD}&=&\frac{N}{2}\sum_{i,j}(F_{i,j}^{\rm A}-F_{i,j}^{{\rm A}*})
                               \varepsilon_i\varepsilon'_j \nonumber \\
                         &=&{\rm i}N\{ 
     (\varepsilon_x\varepsilon'_y-\varepsilon_y\varepsilon'_x){\rm Im} F^{\rm A}_{x,y} 
  + (\varepsilon_y\varepsilon'_z-\varepsilon_z\varepsilon'_y){\rm Im} F^{\rm A}_{y,z} 
        \nonumber \\
 &~&~~+ (\varepsilon_z\varepsilon'_x-\varepsilon_x\varepsilon'_z)
                           {\rm Im} F^{\rm A}_{z,x} \},
\end{eqnarray}
where N denotes the number of the magnetic ion sites.
${\rm Im} F^{\rm A}_{x,y}$  can be written  as
\[ 
       {\rm Im} F^{\rm A}_{x,y}=\frac{1}{2{\rm i}}(F^{\rm A}_{+1}-F^{\rm A}_{-1}),
\]
where $F^{\rm A}_{\pm 1}$
are integrated transition probabilities 
of the $1s \rightarrow 4p$ dipole transition 
with plus and minus helicity light injected
along the $z$-axis for an ion site belongs to the sublattice A. 
Carra {\it et al.} have proposed a sum rule for 
the core-level x-ray absorption spectra with the circularly polarized light\cite{Carra}.
From the sum rule, it is easy to find that the quantity  
$F^{\rm A}_{+1}-F^{\rm A}_{-1}$ is proportional to the z-component of the orbital moment 
of the 4$p$ orbital $\langle L^{4p}_z \rangle$ for a magnetic ion site belongs to the sublattice A in the ground state.
Similarly, the quantities ${\rm Im} F^{\rm A}_{y,z}$ and ${\rm Im}F^{\rm A}_{z,x}$
are proportional to the x- and y-components  of the 4$p$ orbital angular momenta
$\langle L^{4p}_x \rangle$ and $\langle L^{4p}_y \rangle$, respectively.
The integrated amplitude $f_{\rm DD}$ is, therefore, proportional to
$(\mib{\varepsilon} \times \mib{\varepsilon'})\cdot \langle \mib{L}^{4p} \rangle$.

In Fig.~\ref{crymag}, the magnetic and crystal structures in the AFI phase are
shown.  The black and white circles are the V ion sites and the arrows indicate
 the direction of their magnetic moments.
There are eight different V ion sites in the monoclinic unit cell shown in Fig.~\ref{crymag}(a).
In (b) and (c), two different $(010)$ magnetic layers are separately drawn,
and the eight V ion sites are labeled A to H.
The integrated scattering amplitude $f_{\rm DD}(h,k,l)$ of RXS at a 
$(h,k,l)$ Bragg reflection  can be  written in the
form:
\begin{eqnarray}
f_{\rm DD}(h,k,l)&=&\frac{N}{8}\sum_{i,j}\varepsilon_i \varepsilon'_j \bigl[\,
              \{F^{\rm A}_{i,j}+(-)^{h+k+l}F^{\rm F}_{i,j} \} \nonumber\\
           &+& \{F^{\rm B}_{i,j}+(-)^{h+k+l}F^{\rm E}_{i,j} \}(-)^h \nonumber\\
           &+& \{F^{\rm C}_{i,j}+(-)^{h+k+l}F^{\rm H}_{i,j}\}e^{ {\rm i}\alpha } \nonumber\\
           &+& \{F^{\rm D}_{i,j}+(-)^{h+k+l}F^{\rm G}_{i,j} \}(-)^he^{ {\rm i}\alpha} \,\bigr],
\end{eqnarray}
where $\alpha\sim -4\pi(hx+lz)$. $x\sim$0.3438 and $z\sim$0.2991 are the position parameters of the V ions.\cite{Dernier}
If we assume the antiferromagnetic structure above, the system 
is invariant under the symmetry operation 
$\{\Theta| \frac{1}{2}(\mib{a}_m+\mib{b}_m+\mib{c}_m) \}$ and by this operation, the
V ion sites A, B, C and D are transformed  into the sites F, E, H and G, respectively 
and as we have discussed above, relations $F^{\rm A}=F^{{\rm F} *}$, %
$F^{\rm B}=F^{{\rm E} *}$, $F^{\rm C}=F^{{\rm H} *}$ and $F^{\rm D}=F^{{\rm G} *}$
are satisfied.
Similarly, the V ion sites A, C, E and G are transformed  into the sites B, D,  F and H
by the $I2/a$ space group symmetry operation 
$\{\sigma_{\rm h}| \frac{1}{2}\mib{a}_m \}$
and relations $F^{\rm A}=F^{\rm B}$, %
$F^{\rm C}=F^{\rm D}$, $F^{\rm E}=F^{\rm F}$ and $F^{\rm G}=F^{\rm H}$
are obtained.
Using above relations among  $F^l$'s and also the fact that
the operation $\sigma_{\rm h}$ transforms
$(\varepsilon_x,\varepsilon_y,\varepsilon_z)$ into 
$(\varepsilon_x,\varepsilon_y,-\varepsilon_z)$, we obtain the scattering amplitude
for the reflections with $h=$even and $h+k+l=$odd as
\begin{eqnarray}
       &~&f_{\rm DD}(h,k,l)= \nonumber \\
       &~&~~~~\frac{{\rm i}N}{2}\{ (\varepsilon_y\varepsilon'_z-\varepsilon_z\varepsilon'_y)
                ({\rm Im}F^{\rm A}_{y,z}+ e^{{\rm i}\alpha}{\rm Im}F^{\rm C}_{y,z}) \nonumber \\
       &~&~~~ + (\varepsilon_z\varepsilon'_x-\varepsilon_x\varepsilon'_z) 
                 ({\rm Im}F^{\rm A}_{z,x}+ e^{{\rm i}\alpha}{\rm Im}F^{\rm C}_{z,x})\}.\label{f221} 
\end{eqnarray}
On the other hand, the scattering amplitude for the reflections with $h=$odd and $h+k+l=$odd 
can be written as
\begin{eqnarray}
       &~&f_{\rm DD}(h,k,l)= \nonumber \\
      &~&~~~~\frac{{\rm i}N}{2} (\varepsilon_y\varepsilon'_z-\varepsilon_z\varepsilon'_y)
                ({\rm Im} F^{\rm A}_{y,z}+ e^{{\rm i}\alpha} {\rm Im}  F^{\rm C}_{y,z}).\label{f111}
\end{eqnarray}

It is obvious from eqs.~(\ref{f221}) and (\ref{f111}) that the amplitudes are zero
for the $\sigma$-$\sigma$ scattering, where the polarization
vectors $\mib{\varepsilon}$ and $\mib{\varepsilon'}$ are perpendicular
 to the scattering plane and thus are parallel to each other.
This is consistent with the experiments, where only vanishingly small intensities 
of both the $(111)$ and $(221)$ reflections for the $\sigma$-$\sigma$ scattering 
are observed\cite{Paolasini}.
On the other hand,  the $\sigma$-$\pi$ scattering, 
where one of the polarization vectors is perpendicular
to the scattering plane and the other is in the plane, is not forbidden scattering
 and indeed, for the magnetic $(221)$ Bragg reflection, strong RXS are observed in the experiment.
The experimental $\sigma$-$\pi$ scattering at the $(111)$ 
Bragg reflection, however, has only negligible small intensity in the $1s \rightarrow 4p$ 
excitation energy region.
This fact can be explain as follows.
Since the direction of the magnetic moment observed in the neutron experiment
 is almost in the $\mib{a}_m$-$\mib{b}_m$ plane, the x-component of the
 4$p$ orbital angular momentum $\langle L^{4p}_x \rangle$,
 which is perpendicular to the $\mib{a}_m$-$\mib{b}_m$ 
 plane (see Fig.~\ref{crymag}), is expected to be zero or small.
Because eq.~(\ref{f111}) only contains $F^{\rm A}_{y,z}$ and $F^{\rm B}_{y,z}$,
 which is proportional to $\langle L^{4p}_x \rangle$, the intensity should be almost zero.
 
\subsection{Interference of $1s \rightarrow 4p$ and $1s \rightarrow 3d$ processes}
We have discussed the $1s \rightarrow 4p$ dipole RXS.
Only the dipole process, the scattering amplitude at the $(111)$ Bragg reflection should be 
almost zero and thus the $1s \rightarrow 3d$ quadrupole transition expected 
to be important to explain the sharp peak observed at the prethreshold region.
Since V$^{3+}$ ion sites do not have inversion symmetry in V$_2$O$_3$,
the $4p$ and $3d$ orbitals on the same site can be hybridized in each other
and this causes interference between
 the $1s \rightarrow 4p$ dipole transition  and $1s \rightarrow 3d$ quadrupole
transition  processes\cite{Fabrizio}.
Because pure $1s \to 3d$ scattering amplitude is expected to be much smaller
than the interference scattering amplitude, here, only this interference scattering process 
is considered.

The  crystal field which causes
the hybridization between the 4$p$ and 3$d$ orbitals in the same V ion site
would be written as
\begin{eqnarray}
V^l(\mib{r})=&\pm& A_{1,0}r{\rm Y}_{1,0}(\Omega)
                      \pm A_{3,0}r^3{\rm Y}_{3,0}(\Omega)\nonumber\\
      &\pm& A_{3,3}r^3\{{\rm Y}_{3,3}(\Omega)-{\rm Y}_{3,-3}(\Omega)\},
\end{eqnarray}
where the signs in front of three coefficients $A_{1,0}$,  $A_{3,0}$ and $A_{3,3}$ 
depend on the sites:  the plus signs in front of 
$A_{1,0}$ and $A_{3,0}$ are taken for A, B, E and F sites and the minus signs for the other sites
and $+A_{3,3}$ for B, C, E and H sites and $-A_{3,3}$ for the others.

The relative values of the these three parameters 
were estimated using a VO$_6$ cluster model and they are
$A_{3,3}/A_{3,0}\simeq 0.79$ and
$(A_{3,0}\langle 4p|r^3|3d \rangle)/(A_{1,0}\langle 4p|r|3d \rangle)\simeq -0.22$.
If we take into account this hybridization as a second order perturbation with 
respect to $V^l$,
we find for the interference term $f_{\rm DQ}(\omega)$ as
\begin{eqnarray}
 &~&f_{\rm DQ}(\omega) ={\rm i}q^2 \sum_l e^{ {\rm i}(\mib{q}-\mib{q'})\cdot \mib{R}_l } 
 \nonumber\\
 &~&~~\times \Bigl[ \langle{\rm  g} |  
           D( \mib{ \varepsilon' } ) (\frac{V^l}{\Delta_{4p\mbox{-}3d} })
           \frac{1}{ E_g+\omega-H_{\rm m}+{\rm i}\frac{\Gamma_{\rm m}}{2 }}
           Q( \mib{ \varepsilon }, \mib{ q } )       
     | {\rm g} \rangle \nonumber\\
&~&~~~~- \langle{\rm  g} |  
          Q( \mib{ \varepsilon' }, \mib{ q' } )  
           \frac{1}{ E_g+\omega-H_{\rm m}+{\rm i}\frac{\Gamma_{\rm m}}{2} }
          (\frac{V^l}{\Delta_{4p\mbox{-}3d} })  D( \mib{ \varepsilon } ) 
       | {\rm g} \rangle \Bigr],\nonumber\\
\end{eqnarray}
where $| {\rm g} \rangle$ and $E_g$  are the ground state wavefunction and its energy,
$D( \mib{ \varepsilon} )$ and $Q( \mib{ \varepsilon}, \mib{ q } )$ are the operators
for the electric dipole and quadrupole transitions 
defined by  $D( \mib{ \varepsilon } )=e(\mib{ \varepsilon } \cdot \mib{r})$
and $Q( \mib{ \varepsilon }, \mib{ q } )=e( \mib{r}\cdot \mib{ \varepsilon })(\mib{r}\cdot \mib{q})$, $\Delta_{4p\mbox{-}3d}$ is the energy difference
between the $4p$ and $3d$ levels and $H_{\rm m}$ and $\Gamma_{\rm m}$ are
the Hamiltonian and the life-time term for the intermediate state reached
by the $1s \to 3d$ excitation from the initial state.

The term
$1/(E_g+\omega-H_{\rm m}+{\rm i}\Gamma_{\rm m}/2)$
would be approximately replaced by
 $-{\rm i}\pi\sum_\mu \delta(E_g+\omega-E_\mu)|\mu\rangle\langle \mu|$,
where $H_{\rm m}|\mu\rangle=E_\mu|\mu\rangle$ and
with this approximation,
the RXS amplitude integrated by $\omega$ over 
 $1s \rightarrow 3d$ excitation energy region
can be written as
 \begin{eqnarray}
 &~&\int_{1s \rightarrow 3d}  f_{\rm DQ}(\omega)\,d\omega \nonumber\\
 &\sim&\pi(eq)^2\frac{\langle 1s|r|4p\rangle\langle 1s|r^2|3d\rangle}{\Delta_{4p\mbox{-}3d} }\nonumber\\
 &\times& \sum_{l,m,m'} e^{ {\rm i}(\mib{q}-\mib{q'})\cdot \mib{R}_l } 
     \langle g| \sum_\sigma a_{lm\sigma} {a^\dagger}_{lm'\sigma} |g \rangle \nonumber\\
 &\times&\Bigl[D^*_m(\mib{\varepsilon'})V^l_{m,m'}Q_{m'}( \mib{ \varepsilon }, \mib{ q } ) 
  -Q^*_m( \mib{ \varepsilon' }, \mib{ q' } )V^{l*}_{m',m}D_{m'}(\mib{\varepsilon}) \Bigr],\nonumber\\ 
  \label{fDVQ}
\end{eqnarray}
where  $V^l_{m,m'}=\int R_{4p}(r)Y^{*}_{1m}(\Omega)V^l(\mib{r})R_{3d}(r)Y_{2m'}(\Omega)\,d\mib{r}$
and $a^\dagger_{lm\sigma}$ denotes creation operator of an electron on the
3$d$ orbital with magnetic $m$ and spin $\sigma$ quantum numbers on site $l$.
 $D_m$ and $Q_m$ are coefficients of the spherical expansions of the transition operators\label{IntfDVQ}
\begin{eqnarray*}
     D( \mib{ \varepsilon } )&=&er\sum_{m=-1}^1 D_m(\mib{ \varepsilon })Y_{1m}(\Omega)\\
      Q( \mib{ \varepsilon }, \mib{ q } ) &=&
er^2\sum_{m=-2}^{2} Q_m( \mib{ \varepsilon }, \mib{ q } )Y_{2m}(\Omega)
\end{eqnarray*}
and they are given by
\begin{eqnarray*}
    D_{\pm 1} &=& \sqrt{\frac{4\pi}{3}} \frac{ \mp \varepsilon_x + {\rm i} \varepsilon_y }{ \sqrt{2} },
  ~~~~~~~~~D_0   = \sqrt{\frac{4\pi}{3}} \varepsilon_z , \\
    Q_{\pm 2}&=& \sqrt{ \frac{2\pi}{15} } 
                 \bigl\{       (q_x\varepsilon_x - q_y\varepsilon_y)
                           \mp {\rm i}(q_x\varepsilon_y + q_y\varepsilon_x)  \bigr\}, \\
     Q_{\pm 1}&=& \sqrt{ \frac{2\pi}{15} } 
                 \bigl\{  \mp(q_z\varepsilon_x + q_x\varepsilon_z)
                           + {\rm i}(q_y\varepsilon_z + q_z\varepsilon_y)  \bigr\},\\
      Q_0           &=& \sqrt{ \frac{4\pi}{5} } q_z\varepsilon_z. 
\end{eqnarray*}

The procedure of the calculations is following.
Eight V ion sites (A to H in Fig.~\ref{crymag}) in the monoclinic unit cell was considered to be four
independent V ion pairs. 
For each of them, the ground state wavefunction was calculated using the V ion pair model
described in \S~\ref{Model}.
 In the calculations, molecular field  $H_{\rm M}=0.01$~eV
along the magnetic easy axis
was applied so to reproduce experimentally observed antiferromagnetic order.
Then, from the ground state wavefunctions, the quantity
  $\langle g| \sum_\sigma a_{lm\sigma} {a^\dagger}_{lm'\sigma}|g \rangle$
 were calculated, 
 and using  eq.~(\ref{fDVQ}), the scattering amplitude was obtained.

\subsection{Analysis of the (111) Bragg reflection\label{ANA111}}

\begin{figure}
\begin{center}
%\figureheight{3cm}
\epsfig{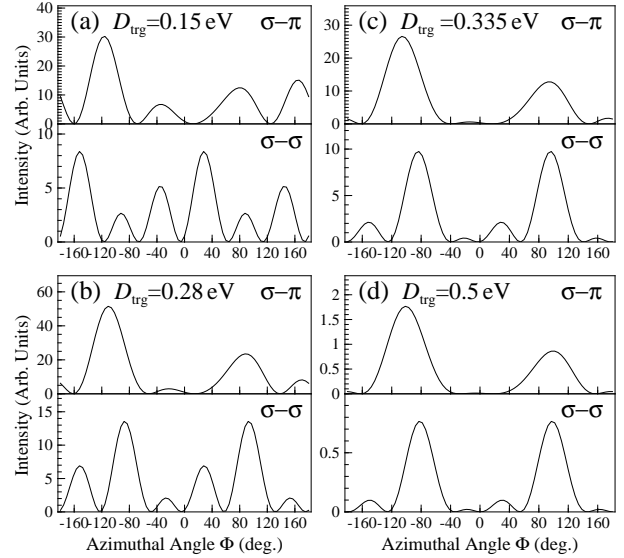}
\end{center}\par
\caption{
 Azimuthal angle dependence of the 
 integrated intensities for the  $(111)$ Bragg reflection in the AFI phase
 calculated with various values of the trigonal crystal field $D_{\rm trg}$.
 Intensities for the two porlarization conditions 
$\sigma$-$\sigma$ and $\sigma$-$\pi$ are shown.
The adopted parameter values are $D_{\rm trg}=0.15$~eV (a), 
$D_{\rm trg}=0.28$~eV (b), $D_{\rm trg}=0.335$~eV (c) and $D_{\rm trg}=0.5$~eV (d).\label{RXS}
} 
\end{figure}
Figure \ref{RXS} shows the theoretical azimuthal angle $\Phi$ dependence of the integrated
intensities for the $(111)$ Bragg reflection in the AFI phase
for  various values of $D_{\rm trg}$. The intensities with the two polarization conditions 
$\sigma$-$\sigma$ and $\sigma$-$\pi$ calculated with the parameters 
$D_{\rm trg}=0.15$~eV (a), $D_{\rm trg}=0.28$~eV (b), $D_{\rm trg}=0.335$~eV  (c)  and $D_{\rm trg}=0.5$~eV (d) are shown. The scattering plane is parallel to the $z$-axis at $\Phi=0$.
The tilting angle of the magnetic moment $\theta_{\rm M}$ in the ground state
with $D_{\rm trg}=0.15$~eV, $0.28$~eV, $0.335$~eV and $0.5$~eV are
$\theta_{\rm M}=16^\circ$, 51$^\circ$, 71$^\circ$ and 84$^\circ$, respectively.

It is seen in the figure that the azimuthal angle $\Phi$ dependence
of the scattering intensities are strongly influence by the direction 
of the orbital magnetic moment.
The intensity curve in Fig.~\ref{RXS}(c) with $\theta_{\rm M}=71^\circ$, which is the same
angle observed in the neutron diffraction experiment\cite{Moon} and Fig.~\ref{RXS}(b) 
with $\theta_{\rm M}=51^\circ$ are most resemble to the experimental intensity
curve in ref.~\citen{Paolasini}. On the other hand, the curve in Fig.~\ref{RXS}(b), where $\theta_{\rm M}=16^\circ$ and the ground state is
 almost pure $e^\pi a_1; e^\pi e^\pi$ ($e^\pi e^\pi; e^\pi a_1$) state,
is very different from experimental one.
This indicates that the experimental results are well explained with 
the $e^\pi e^\pi;e^\pi e^\pi$ and $e^\pi a_1;e^\pi e^\pi$ mixed ground state and
consistent with the present theory.
It is also important to mention that if  the spin-orbit interaction is switched
off, the scattering intensity is zero for this reflection, although there
exists the molecular ferro-orbital order in the ground state.
The reflection, therefore, originates from not the molecular ferro-orbital
order but the AF magnetic order.
\begin{figure}
\begin{center}
%\figureheight{3cm}
\epsfig{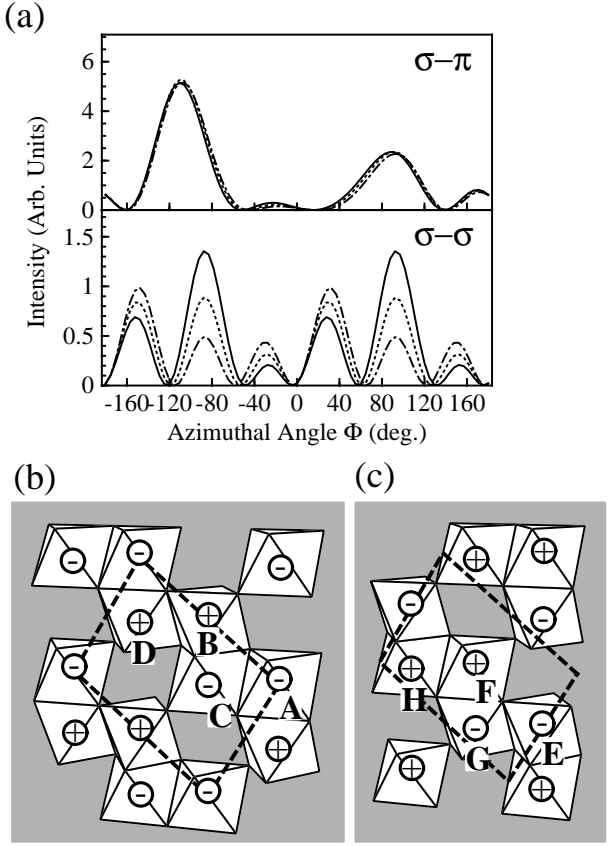}
\end{center}\par
\caption{In (a), azimuthal angle dependence of the 
 integrated intensities for the  $(111)$ reflection in the AFI phase
 calculated with $C_{\rm A}(e_u^\pi,e_v^\pi)=-0.02$~eV (solid line), 0.0~eV(dashed line) 
 and 0.02~eV (dotted-dashed line) are shown.
 In (b) and (c),  the same as Fig.~\ref{crymag} (c) and (d),
 the sign ($\pm$) of $M_{Lx}$ on V ions are shown.}
 \label{magLx}
\end{figure}

The scattering intensity is also 
sensitive to a small x-component of the orbital magnetic moment $M_{Lx}$.
Possibility of the presence of such small x-component of the magnetic moment
was pointed out by Moon\cite{Moon}.
 $M_{Lx}$ can be induced by the low symmetry crystal field $C_{\rm A}(e_u^\pi,e_v^\pi)$.
In Fig.~\ref{magLx}(a),  $\Phi$ dependence of the integrated
intensities for the $(111)$ reflection obtained with $C_{\rm A}(e_u^\pi,e_v^\pi)=-0.02$~eV (solid line), 0~eV (dashed line) and 0.02~eV (dotted-dashed line) are drawn. $D_{\rm trg}=0.28$~eV
is adopted.
In Figs.~\ref{magLx}(b) and (c), the sign of $M_{Lx}$ on each V ion for $C_{\rm A}(e_u^\pi,e_v^\pi)<0$ is shown for the two magnetic layers in Figs.~\ref{crymag}(b) and (c).
For $C_{\rm A}(e_u^\pi,e_v^\pi)>0$, all the signs of $M_{Lx}$ are altered.
As seen in the Fig.~\ref{magLx}(a), the  $\sigma$-$\sigma$ scattering intensity is strongly 
influenced by the value of $C_{\rm A}(e_u^\pi,e_v^\pi)$, although the magnitude of $M_{Lx}$ induced by $C_{\rm A}(e_u^\pi,e_v^\pi)=\pm 0.02$~eV is an order of $\sim 0.01\mu_{\rm B}$. 
With $C_{\rm A}(e_u^\pi,e_v^\pi)=0.02$~eV the calculated
intensity curve is not resemble to the experimental one
whereas with $C_{\rm A}(e_u^\pi,e_v^\pi)=-0.02$~eV, good agreement with experimental 
intensity curve is achieved.
\begin{figure}
\begin{center}
%\figureheight{3cm}
\epsfig{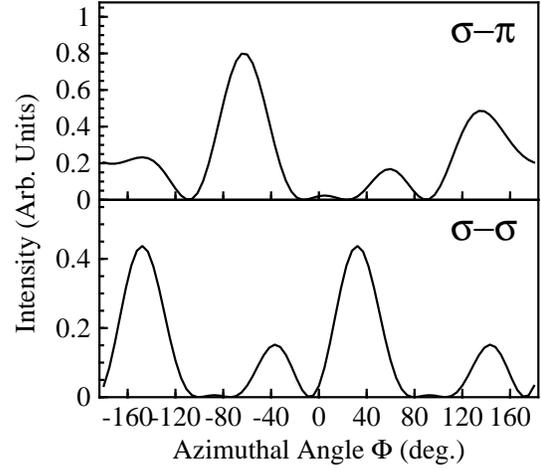}
\end{center}\par
\caption{Azimuthal angle dependence of the 
 integrated intensity for the  $(111)$ reflection in the AFI phase
 calculated with $D_{\rm trg}=0.28$~eV.  The $1s \to 3d$ quadrupole transition instead
 of the $1s \to 3d$  and $1s \to 4p$ interference process 
 is assumed as for the RXS process.
  \label{rxsqqang}
}
\end{figure}

Lovesey and Knight argued that the $(111)$ reflection peak can be 
explain by pure $1s \to 3d$ quadrupole transition\cite{Lovesey}.
Contrary to this, in the present theory, the peak is attributed to 
 the interference process of the $1s \to 4p$ dipole 
and $1s \to 3d$ quadrupole transitions.
 For comparison, in Fig.~\ref{rxsqqang}, $\Phi$ dependence
of the scattering intensity of the $(111)$ reflection peak
obtained with the pure $1s \to 3d$ quadrupole transition instead
of the interference process is shown. The intensity curve
calculated with $D_{\rm trg}=0.28$~eV is very different from experimental curve. 
Results obtained with other parameter values of $D_{\rm trg}$ and $C_{\rm A}(e_u^\pi,e_v^\pi)$
also does not agree with experimental intensity curve.

\section{Linear Dichroism in V 2$p$ XAS}

In this section, the dependence of the V 2$p$ XAS spectra
on the polarization vector $\mib{\varepsilon}$ of the light will be discussed.
As was mentioned in the previous sections, such experiment 
was already done by Park {\it et al.}\cite{Park} with the polarization 
vectors $\mib{\varepsilon}$ being parallel and perpendicular to the $\mib{c}_h$-axis.
Because of the selection rule of the $2p \to 3d$ dipole transition, 
the transition probability to the $a_1$ orbital is larger than
that to the $e^\pi$ orbital for $\mib{\varepsilon} // \mib{c}_h$, whereas
that to the $e^\pi$ orbital is  larger  for the $\mib{\varepsilon} \perp \mib{c}_h$. 
Thus the difference spectrum is sensitive to the $e^\pi a_1$ to $e^\pi e^\pi$
occupation ratio in the initial state.
The ratio was deduced from the experiment using a VO$_6$ cluster
model and was found to be $e^\pi a_1$: $e^\pi e^\pi\sim 1:1$ for the PM phase
and $e^\pi a_1: e^\pi e^\pi\sim 1:2$ for the AFI phase.

To verify the present theory, the spectra was calculated using the V ion pair model described
in \S~\ref{model}.
For the initial state, the ground state of the  
V ion pair model is assumed.
For the final state, the configurations
$2\underline{p}3d^3;3d^2$, $2\underline{p}3d^1;3d^3$ and $2\underline{p}3d^4;3d^1$
are considered, where 2$\underline{p}$ denotes a hole on the V 2$p$ orbital in
one of the V ion site.
In addition to the interactions 
considered in the Hamiltonian described in \S~\ref{model},
the 3$d$-2$p$ multipole interaction and the 
2$p$ spin-orbit interaction are also included in the calculation
of the final states\cite{Tanaka3}.
The $2p \to 3d$ dipole transition is considered as
the photoprocess
and the intensities of the spectra $I_\kappa(\omega)$ as a function
of the incident photon energy $\omega$ is written as
\begin{eqnarray}
  I_\kappa(\omega)&\propto&\sum_f \big|\langle f|T_\kappa|g\rangle\big|^2 
  \delta(E_g+\omega-E_f) \nonumber\\
  &=&\lim_{\delta \to 0} \frac{-1}{\pi}{\rm Im}\langle g| T^\dagger_\kappa\frac{1}{E_g+\omega-H_f+{\rm i}\delta}T_\kappa|g \rangle,
\end{eqnarray}
where $|f \rangle$, $E_f$ and $H_f$ denote the wavefunction, eigen energy and
Hamiltonian of the final state, respectively, $T_\kappa$ represents the $2p \to 3d$ 
dipole transition operator and $\kappa$ specifies the polarization 
direction of the linearly polarized light. 
The spectra was calculated using the recursion method\cite{Haydock}
and a Lorentzian 
broadening was further applied the spectra.

The parameters used in the calculations are the same to those
listed in Table \ref{param} and $D_{\rm trg}=0.28$~eV is adopted.
In the final state, there are additional parameters related to the $2p$
core-hole: the
3$d$-2$p$ Coulomb repulsion energy $U_{dp}=3.07$~eV,
the 2$p$ spin-orbit interaction $\zeta_p=4.649$~eV,
the 3$d$-2$p$ Slater integrals $F^2_{dp}=4.361$~eV, $G^1_{dp}=3.162$~eV and 
$G^3_{dp}=1.797$~eV.
\begin{figure}[htbp]
\begin{center}
%\figureheight{3cm}
\epsfig{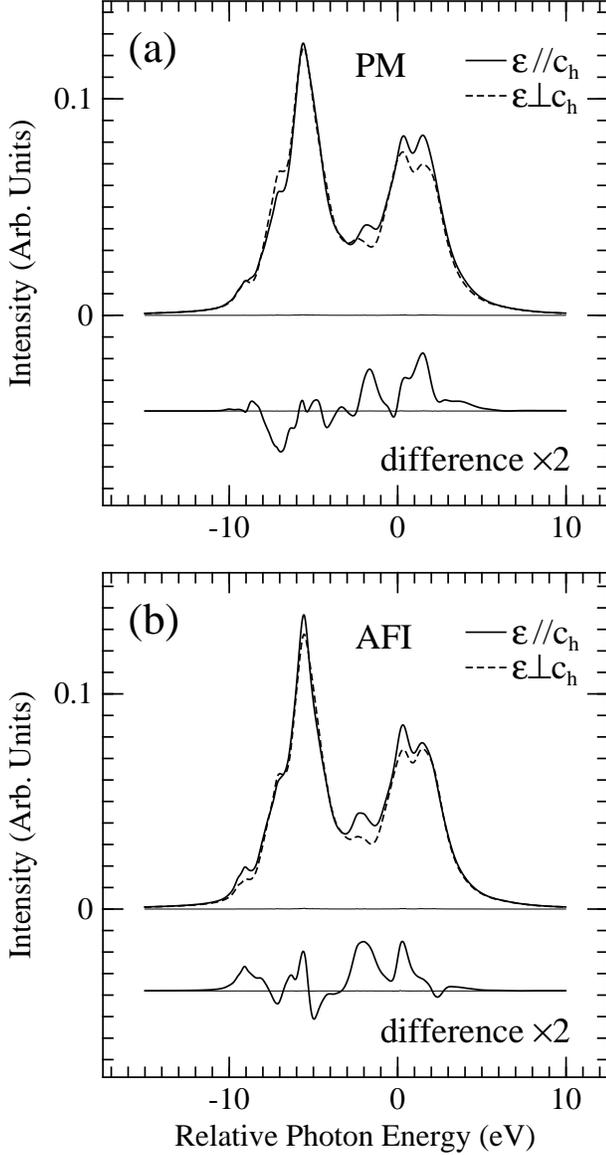}
\end{center}\par
\caption{Theoretical polarization-dependent 
V 2$p$ XAS spectra for V$_2$O$_3$ for the PM (a) and AFI (b) phases. 
The spectra are depicted for two different directions of 
the polarization vectors $\mib{\varepsilon}$ of the light:
parallel and perpendicular to the $\mib{c}_h$-axis.
The difference spectra are also shown below in each panel.
 }\label{xasim}
\end{figure}
\begin{figure}
\begin{center}
%\figureheight{3cm}
\epsfig{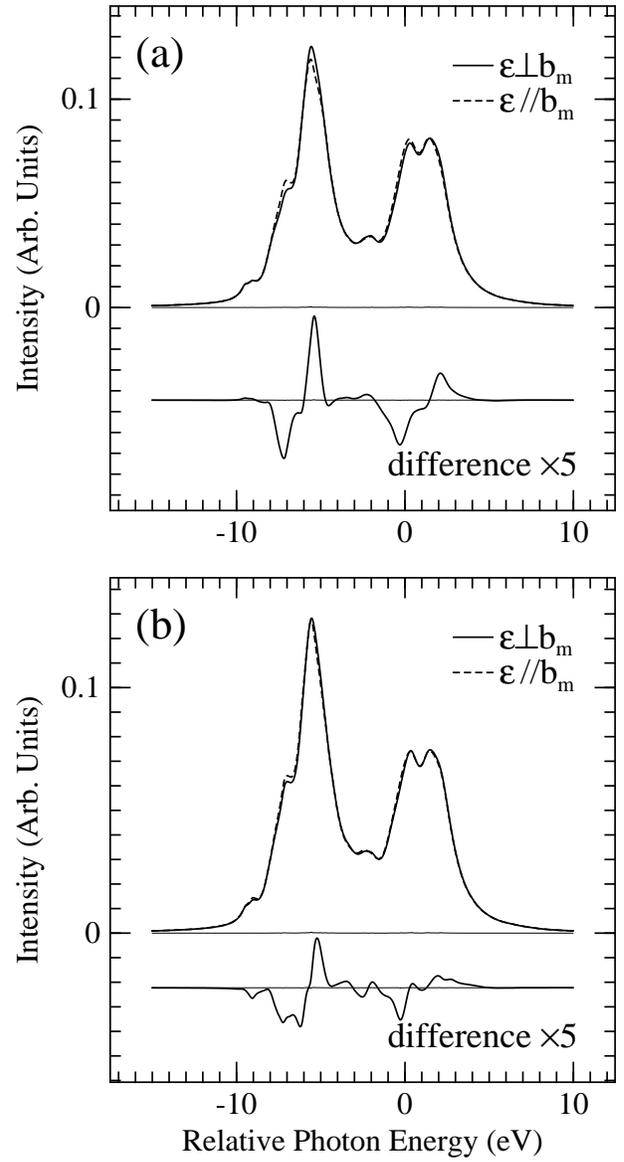}
\end{center}\par
\caption{Theoretical polarization-dependent 
V 2$p$ XAS spectra for the AFI phase calculated without (a)
and with (b) the 3$d$ spin-orbit interaction. 
$D_{\rm trg}=0.28$~eV is assumed.
The spectra are depicted for two different directions of 
the polarization vectors $\mib{\varepsilon}$ of the light:
 parallel and perpendicular to the monoclinic $\mib{b}_m$ axis.
The difference spectra are also shown below in each panel.
}\label{xaso}
\end{figure}
%\clearpage

Figure \ref{xasim} shows the calculated spectra obtained using the V ion
pair model with the polarization vectors $\mib{\varepsilon}$ 
parallel and perpendicular to the $\mib{c}_h$-axis.
In Fig.~\ref{xasim}(a) the spectra for the PM phase, and 
in Fig.~\ref{xasim}(b) those for the AFI phase are
shown and in each panel, the difference spectrum is also depicted.
The theoretical spectra well reproduce experimental ones especially those
for the AFI phase in ref.~\citen{Park}.
 Although the difference spectrum for the PM phase is less similar
to the experimental one, its tendency to have 
 negative peaks at -7~eV and 0~eV
 is consistent with the experiment.
The polarization of the 
spectra defined by $P=(I_{\parallel}-I_{\perp})/(I_{\parallel}+I_{\perp})$ are
$P=0.027$ for the AFI phase and $P=0.015$ for the PM phase, where
$I_{\parallel}$ and $I_{\perp}$ denote integrated intensity of the 
spectra with the light polarization parallel and perpendicular to the $\protect\mib{c}_h$-axis,
respectively. Such large change in the polarization between the
two phases is also consistent with the experiment.
Note that although the electron occupation of the $e^\sigma$ orbitals in the ground state
is small $\sim 5$\%, the inclusion of the $e^\sigma$ orbital is important
to reproduce the spectra, especially its polarization.

The sensitivity of the V 2$p$ XAS spectra with polarized light
to the orbital occupation in the initial state
is useful to ascertain whether the orbital order exists or not
in the AFI phase. 
To demonstrate this, in Fig.~\ref{xaso} the polarization-dependent 
V 2$p$ XAS spectra are shown for two different direction of 
$\mib{\varepsilon}$ in the plane which is perpendicular to the $\mib{c}_h$-axis:
 $\mib{\varepsilon} \perp \mib{b}_m$ and $\mib{\varepsilon}  // \mib{b}_m$.
The spectra without (a) and with (b) the 3$d$ spin-orbit interaction is presented;
other parameters are the same to those adopted for the AFI phase.
The difference spectra are also shown below in each panel.

As was discussed in \S~\ref{GS}, in the absence of the 3$d$ spin-orbit interaction, the
ground state has an orbital polarization 
corresponding to the molecular ferro-orbital ordering
state of Milla's model (Fig.~\ref{xaso}(a)). 
On the other hand, in the presence of the spin-orbit
interaction, the ground state has no orbital ordering and instead of this a
large orbital magnetic moment is induced in the ground state (Fig.~\ref{xaso}(b)).
The magnitude of the difference spectrum in Fig.~\ref{xaso}(a) is much larger
than that in Fig.~\ref{xaso}(b)  and the shape of the spectrum 
in Fig.~\ref{xaso}(a) largely vary from that in Fig.~\ref{xaso}(b).
Thus the measurement of this linear dichroic V 2$p$ spectra,
will be a crucial check whether 
the large orbital moment is present or not.

\section{Conclusions} 
The 3$d$ electronic structure
 and phase transition in pure and Cr doped V$_2$O$_3$ were theoretically 
investigated in relation to the 3$d$ spin-orbit interaction and lattice distortion.
The 3$d$ electronic state of V$_2$O$_3$ have been discussed,
using a V ion pair model on the basis of the configuration interaction
approach. In the model, the 3$d$-3$d$ multipole interaction, 
a trigonal crystal field and the 3$d$ spin-orbit interaction were considered and 
electron hopping between nearest neighbor V ions was also taken into account.
In the presence of the 3$d$ spin-orbit interaction, spin and orbital degrees of freedom
are strongly coupled and this removes the orbital degeneracy of the
ground state, resulting in no orbital ordering.
Instead of this, each V ion with $S=1$ spin state
has a large orbital magnetic moment $\sim 0.7\mu_{\rm B}$ in the AFI phase.
These results are in contrast to those obtained from the 
similar V ion pair model proposed by Mila {\it et al.}, in which 
a ferro molecular orbital ordering occurs in the AFI phase\cite{Mila,Shiina}.
The large orbital moment is in accordance with the ratio 
 $M_{\rm L}/M_{\rm S} \sim -0.3$ deduced from
the nonresonant magnetic x-ray scattering experiment\cite{Paolasini} and
also explains small total magnetic moment $1.2\mu_{\rm B}$ observed in the AFI phase\cite{Moon}.
Because of the presence of the large orbital magnetic moment, the monoclinic lattice 
distortion and the direction of the magnetic easy axis is strongly connected in the AFI phase.
The direction of magnetic moment was found to be canted from the corundum 
$\mib{c}_h$-axis in the presence of the monoclinic distortion
and this agrees with the neutron diffraction experiments\cite{Moon}.

The local lattice distortion energy due to tilting of the V ion pair
 from the corundum $c$-axis was further considered.
From the present model, it is expected that 
there is at least two local minima
for the free energy in the real system as a function of the V-V distance of the pair.
One is positioned at the V-V distance in the PM phase, where the energy difference of the
two kinds of the state $e^\pi a_1;e^\pi e^\pi$ and $e^\pi e^\pi;e^\pi e^\pi$ 
is large and thus no electron-lattice coupling takes place.
The electronic state of the V ion pair in this phase is
described  as a superposition of two configurations 
$e^\pi a_1;e^\pi e^\pi$ and $e^\pi e^\pi;e^\pi a_1$ in equal weight, 
which is similar to Mila's model.
The other one is located at the V-V distance in the PI and AFI phases,
where energies of the two kinds of the states are degenerate
and Jahn-Teller like lattice distortion is
the cause of the energy lowering at this V-V distance.
Because of the tilting of the V ion pair caused by this lattice instability, 
the $e^\pi e^\pi;e^\pi e^\pi$ state is further hybridized
with above $e^\pi a_1;e^\pi e^\pi$ ($e^\pi e^\pi;e^\pi a_1$) state in these phases.
This large difference in the electron occupation between the metal and insulating phases
agrees with the recent linear dichroic V 2$p$ x-ray absorption spectroscopy (XAS) experimental results\cite{Park}. 

Since both in the AFI and PI phases, the electron-lattice coupling strongly influences
the orbital occupation and charge fluctuation,
the PM $\to$ AFI and PM $\to$ PI transitions can not be regarded
as usual Mott transition.
The elongation of the V-V distance and resultant reduction in the hybridization strength
between the 3$d$ orbitals caused by this electron-lattice coupling
is probably responsible for these metal-insulator transitions.

To investigate interplay between magnetic ordering and the lattice distortion, an
effective spin Hamiltonian including the effects of tilting of the V ion pairs was
introduced. 
The antiferromagnetic order with the monoclinic 
lattice distortion observed in the AFI phase was
reproduced in this model, where simultaneous ordering of the tilting directions 
and the magnetic moments of the V ion pairs take place.
With increasing temperature, it exhibits antiferromagnetic
to paramagnetic phase transition corresponding to the AFI $\to$ PI 
transition in the real system.
While all V ion pairs are tilted to the same direction 
and this gives rise the monoclinic lattice distortion in the AFI phase,
their tilting directions are disordered and fluctuate among the three stable directions in the PI phase. As a result, there is no monoclinic distortion in the PI phase.
It was also found that the $C_{44}$ elastic constant  
softens and short-range spin correlation functions abruptly changes 
at this second order transition in this model.
These results are consistent with the ultrasonic-wave measurement\cite{Yang}
and recent neutron scattering experiments\cite{Bao,Bao2}.

To confirm the validity of the present model, recent experiments on 
the resonant X-ray scattering (RXS) at the V  $K$-edge\cite{Paolasini,Paolasini2}
and the linear dichroic V 2$p$ XAS experiments\cite{Park}
were analyzed.
The $(111)$ Bragg reflection observed in the RXS experiments, particularly in 
its azimuthal angle and polarization dependence of the $1s \to 3d$ peak, is well explained within
 the present model, where no orbital ordering is present and the antiferromagnetic
 order is assumed. The reflection is pure magnetic and 
the scattering amplitude of this reflection mainly arises from the
 interference process of the $1s \to 4p$ dipole and  $1s \to 3d$ quadrupole transitions.
The shape and magnitude of the linear dichroic V 2$p$ XAS spectra
in both the AFI and PM phases are well reproduced within the present model.
Another linear dichroic V 2$p$ XAS experiment has been proposed,
which is sensitive to the presence of the orbital order in the initial state
and will be a crucial check to the present model.

This work is partly supported by a Grant-in-Aid for Scientific Research from the Ministry
of Education, Culture, Sports, Science and Technology.

\end{document}